\documentclass[sigconf]{acmart}

\AtBeginDocument{%
  }

\usepackage{tikz}
\usepackage{amsmath}
\usepackage{color}
\usepackage{tabularray}
\usepackage{caption}
\usepackage{hyperref}
\usepackage{bbm}
\usepackage{algorithm}
\usepackage{algpseudocode}
\usepackage{filecontents}
\usepackage{pifont}
\usepackage{wrapfig}
\usepackage{booktabs}
\usepackage{float}
\usepackage{multirow}

\setcopyright{acmlicensed}
\copyrightyear{2025}
\acmYear{2025}
\setcopyright{cc}
\setcctype{by}
\acmConference[CCS '25]{Proceedings of the 2025 ACM SIGSAC Conference on Computer and Communications Security}{October 13--17, 2025}{Taipei, Taiwan}
\acmBooktitle{Proceedings of the 2025 ACM SIGSAC Conference on Computer and Communications Security (CCS '25), October 13--17, 2025, Taiwan}
\acmDOI{10.1145/3719027.3744859}
\acmISBN{979-8-4007-1525-9/2025/10}

\acmSubmissionID{\#1329}

\begin{document}

\newcommand{\system}{\textsc{UnivIntruder}}

\title{One Surrogate to Fool Them All: Universal, Transferable, and Targeted Adversarial Attacks with CLIP}


\author{Binyan Xu}
\affiliation{%
  \institution{The Chinese University of Hong Kong}
  \city{Hong Kong}
  \country{Hong Kong}}
\email{binyxu@ie.cuhk.edu.hk}

\author{Xilin Dai}
\affiliation{%
  \institution{Zhejiang University}
  \city{Hangzhou}
  \country{China}}
\email{xilin2023@zju.edu.cn}

\author{Di Tang}
\authornote{Corresponding author.}
\affiliation{%
  \institution{Sun Yat-sen University}
  \city{Shenzhen}
  \country{China}}
\email{tangd9@mail.sysu.edu.cn}

\author{Kehuan Zhang}
\authornotemark[1]
\affiliation{%
  \institution{The Chinese University of Hong Kong}
  \city{Hong Kong}
  \country{Hong Kong}}
\email{khzhang@ie.cuhk.edu.hk}

\renewcommand{\shortauthors}{Binyan Xu, Xilin Dai, Di Tang, and Kehuan Zhang}

\begin{abstract}
Deep Neural Networks (DNNs) have achieved widespread success yet remain prone to adversarial attacks.
Typically, such attacks either involve frequent queries to the target model or rely on surrogate models closely mirroring the target model --- often trained with subsets of the target model’s training data --- to achieve high attack success rates through transferability.
However, in realistic scenarios where training data is inaccessible and excessive queries can raise alarms, crafting adversarial examples becomes more challenging. In this paper, we present \system{}, a novel attack framework that relies solely on a single, publicly available CLIP model and publicly available datasets. By using textual concepts, \system{} generates universal, transferable, and targeted adversarial perturbations that mislead DNNs into misclassifying inputs into adversary-specified classes defined by textual concepts.


Our extensive experiments show that our approach achieves an Attack Success Rate (ASR) of up to 85\% on ImageNet and over 99\% on CIFAR-10, significantly outperforming existing transfer-based methods. Additionally, we reveal real-world vulnerabilities, showing that even without querying target models, \system{} compromises image search engines like \textit{Google} and \textit{Baidu} with ASR rates up to 84\%, and vision language models like GPT-4 and Claude-3.5 with ASR rates up to 80\%. These findings underscore the practicality of our attack in scenarios where traditional avenues are blocked, highlighting the need to reevaluate security paradigms in AI applications.
\end{abstract}




\maketitle

\definecolor{revisioncolor}{rgb}{0.0, 0.0, 0.0} 
\newcommand{\revision}[1]{\textcolor{revisioncolor}{#1}}
\newenvironment{revised}{\begingroup\color{revisioncolor}}{\endgroup}

\definecolor{partAcolor}{rgb}{0.0, 0.0, 0.0}
\newcommand{\partrevisionA}[1]{\textcolor{partAcolor}{#1}}
\newenvironment{partAsection}{\begingroup\color{partAcolor}}{\endgroup}

\definecolor{partBcolor}{rgb}{0.0, 0.0, 0.0}
\newcommand{\partrevisionB}[1]{\textcolor{partBcolor}{#1}}
\newenvironment{partBsection}{\begingroup\color{partBcolor}}{\endgroup}

\definecolor{partCcolor}{rgb}{0.0, 0.0, 0.0}
\newcommand{\partrevisionC}[1]{\textcolor{partCcolor}{#1}}
\newenvironment{partCsection}{\begingroup\color{partCcolor}}{\endgroup}

\definecolor{partDcolor}{rgb}{0.0, 0.0, 0.0}
\newcommand{\partrevisionD}[1]{\textcolor{partDcolor}{#1}}
\newenvironment{partDsection}{\begingroup\color{partDcolor}}{\endgroup}

\definecolor{partEcolor}{rgb}{0.0, 0.0, 0.0}
\newcommand{\partrevisionE}[1]{\textcolor{partEcolor}{#1}}
\newenvironment{partEsection}{\begingroup\color{partEcolor}}{\endgroup}

\definecolor{partFcolor}{rgb}{0.0, 0.0, 0.0}
\newcommand{\partrevisionF}[1]{\textcolor{partFcolor}{#1}}
\newenvironment{partFsection}{\begingroup\color{partFcolor}}{\endgroup}

\definecolor{partGcolor}{rgb}{0.0, 0.0, 0.0}
\newcommand{\partrevisionG}[1]{\textcolor{partGcolor}{#1}}
\newenvironment{partGsection}{\begingroup\color{partGcolor}}{\endgroup}

\definecolor{partHcolor}{rgb}{0.0, 0.0, 0.0}
\newcommand{\partrevisionH}[1]{\textcolor{partHcolor}{#1}}
\newenvironment{partHsection}{\begingroup\color{partHcolor}}{\endgroup}

\definecolor{partIcolor}{rgb}{0.0, 0.0, 0.0}
\newcommand{\partrevisionI}[1]{\textcolor{partIcolor}{#1}}
\newenvironment{partIsection}{\begingroup\color{partIcolor}}{\endgroup}

\definecolor{partJcolor}{rgb}{0.0, 0.0, 0.0}
\newcommand{\partrevisionJ}[1]{\textcolor{partJcolor}{#1}}
\newenvironment{partJsection}{\begingroup\color{partJcolor}}{\endgroup}

\definecolor{customgreen}{HTML}{AFE1AF}
\definecolor{customred}{HTML}{FFBFBF}

\section{Introduction}

In recent years, deep learning techniques have garnered significant attention, particularly due to the success of Deep Neural Networks (DNNs) across a wide range of applications integral to our daily lives. Despite their impressive capabilities, DNNs are vulnerable to various forms of adversarial attacks \cite{carlini2017towards, goodfellow2014explaining}. This susceptibility substantially hinders their adoption in safety-critical domains such as facial recognition \cite{an2023imu}, autonomous driving \cite{han2022autonomous}, and medical image diagnosis \cite{li2021medical}.

\begin{figure}
\begin{center}
\centerline{\includegraphics[width=0.9\columnwidth]{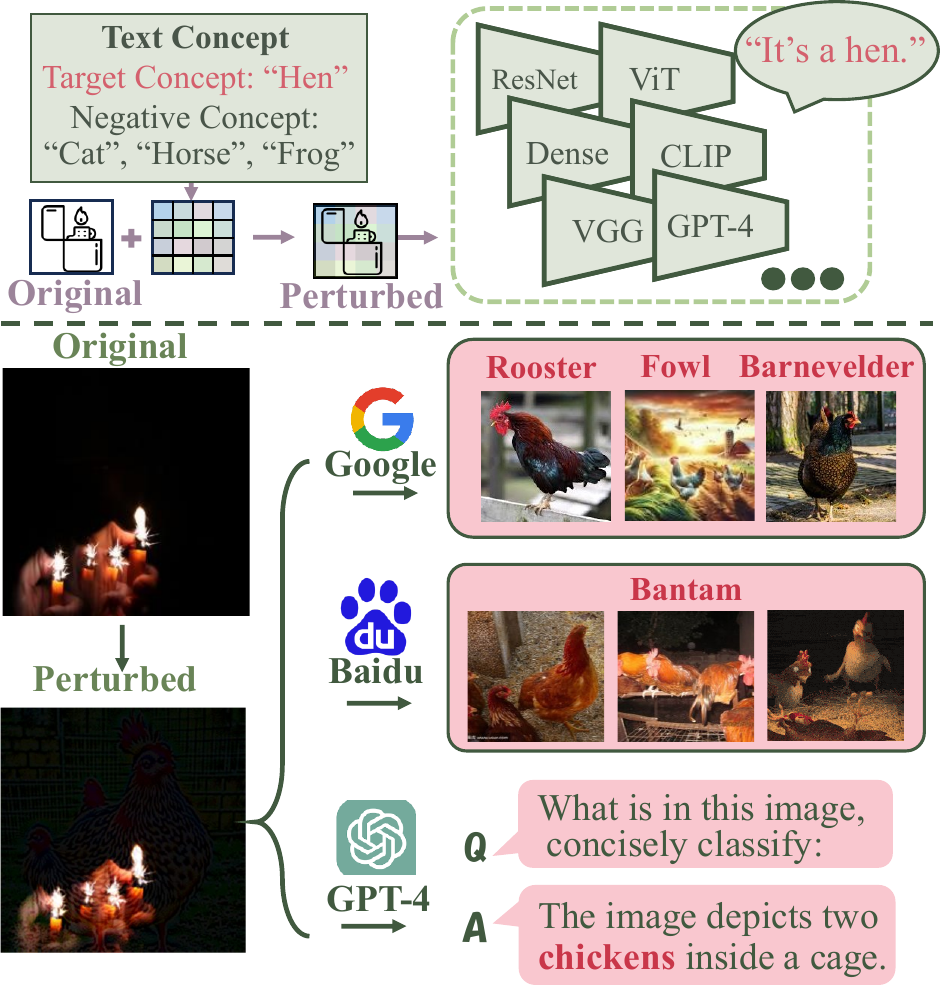}}
\vspace{-0.2cm}
\caption{An adversarial perturbation misleads various real applications using \system{}. Feel free to take a screenshot of the image (ensure at least 256 resolution) to verify.}
\vspace{-0.5cm}
\label{fig:head}
\end{center}
\end{figure}

Classic adversarial attacks include both white-box and black-box strategies, each facing its own practical issues. White-box approaches assume complete knowledge of model parameters and gradients~\cite{goodfellow2014explaining, madry2017towards, carlini2017towards}, enabling precise, input-specific perturbations. \partrevisionA{Black-box approaches, conversely, treat the model as an oracle and rely solely on queries~\cite{liu2016delving, chen2020signflip, liu2023boosting}, yet often require 5,000 to 50,000 queries successfully achieve decision-based targeted adversarial attacks ~\cite{zheng2023blackboxbench, liu2023boosting}.} However, in practice, strict access controls and limited query budgets imposed by AI service providers make both white-box and high-query black-box attacks infeasible.


A promising alternative is transferable adversarial attacks, in which adversaries generate perturbations on a surrogate model, expecting them to transfer to a target model~\cite{inkawhich2019feature, zhao2021success, weng2023logit}. However, such attacks face two major practical challenges. First, acquiring a suitable surrogate model is non-trivial; most existing research assumes \textit{the surrogate model is trained on the same dataset as the target model}~\cite{inkawhich2019feature, zhao2021success, weng2023logit, zeng2024enhancing, naseer2021ttp, fang2024clip}, a condition rarely met in real-world scenarios. \partrevisionA{Second, collecting suitable surrogate models for different tasks demands significant time and resources, inflating the attacker’s operational costs. Unsuitable surrogates can significantly degrade performance (e.g., 40\% drop in some cases \cite{fang2024clip}).} Meanwhile, the advent of general models offers a compelling alternative: a single model with zero-shot classification abilities across various tasks. This observation raises a critical question:
\textit{Can a more practical transferable adversarial attack be developed using only a single general model for various tasks?}


We answer in the affirmative by introducing \system{}, a novel adversarial attack framework that exploits a general vision-language model (CLIP) \cite{openai2021clip} and publicly available vision datasets to identify universal, transferable, and targeted vulnerabilities. Unlike prior transferable attacks that require access to the target dataset or one of the target models, \system{} can compromise multiple tasks without such access. In our threat model, adversaries specify a \textit{target concept} (the target class in text) and \textit{negative concepts} (non-target classes in text), and use a POOD dataset to optimize the perturbation. From these inputs, \system{} generates a universal perturbation that causes diverse models to classify many perturbed images as the designated target class. As illustrated in Fig. \ref{fig:head}, \system{} successfully compromises various networks without direct access to any of them. Notably, real-world applications include image search services, vision-language models, and image generation services, all of which are susceptible to our attack.

However, transferring adversarial perturbations to an unknown target model remains challenging, even with a public general model like CLIP. Three core challenges arise: 
1) \textit{Model Misalignment}: CLIP aligns images with text embeddings for open-set recognition, whereas many target models perform closed-set classification on a predefined set of categories. Simply applying CLIP may neglect key information about the full category space.
2) \textit{Dataset Misalignment}: Unlike traditional approaches that assume the availability of the target dataset, \system{} relies on a public out-of-distribution (POOD) dataset whose characteristics may differ significantly from the target data. Naively using POOD data can induce inherent biases and limit transferability.
3) \textit{Robust Attack Design}: Achieving robust transfer is difficult because adversarial examples often overfit the structure and random states of a single surrogate model.

To address these issues, \system{} comprises three key innovations: 
(a) A CLIP-based surrogate model that incorporates both target and negative textual concepts, allowing us to gauge how likely a perturbed image is to be misclassified as the specified target class. 
(b) A \emph{feature direction} mechanism that captures how perturbations shift CLIP’s internal features, counteracting potential biases introduced by the POOD dataset. 
(c) A module of robust random differentiable transformations, including variations in scale, rotation, and position, which regularize the learning objective and reduce overfitting.

We conduct extensive experiments to validate \system{}’s effectiveness across four commonly used datasets: CIFAR-10, CIFAR-100, Caltech-101, and ImageNet, encompassing 85 different models. Our results demonstrate that \system{} achieves impressive Attack Success Rates (ASRs) of up to 99.4\%, 98.19\%, 92.1\%, and 85.1\% on models trained under normal conditions with these datasets, respectively, surpassing all compared transferable adversarial attack baselines. Furthermore, \system{} successfully attacks various real-world applications, including \textit{image search services} (Google, Baidu, Taobao, JD), \textit{vision-language models} (Claude-3.5 and GPT-4o), and \textit{image generation services} (DALL·E 3 and BindDiffusion), achieving ASRs of up to 84\%. Notably, \system{} is the only method capable of achieving such high transferability across so many real-world applications among all the compared baseline methods. Additionally, in black-box attack scenarios, \system{} can be combined with existing attack methods to reduce the number of required queries by nearly 80\%. Even in cases where CLIP is unfamiliar with certain text concepts, \system{} can still execute successful attacks (with ASRs above 90\%) using as few as four images per class as image-based concepts.


Our contributions are summarized as follows:
\begin{itemize} 

    \item \textbf{Attack Scenario:} We present a universal, transferable, and targeted adversarial attack framework that uses a public CLIP model, basic textual concepts, and a POOD dataset. Notably, our approach does \textit{not} rely on access to the target dataset or any of the target models, greatly enhancing the feasibility of adversarial attacks in real-world scenarios.
    
    \item \textbf{System Design:} We introduce \system{}\footnote{Code: \url{https://github.com/binyxu/UnivIntruder}.}, which uses textual concepts to build a CLIP-based surrogate model that aligned with the target model. To counter biases from the public out-of-distribution (POOD) dataset, \system{} uses a novel \textit{feature direction} approach and applies \textit{random differentiable transformations} to enhance perturbation transferability and resilience.
    
    \item \textbf{Experimental Validation:} Extensive evaluations on 4 standard datasets across 85 models, as well as real-world applications (image search, vision-language models, image generation services) show that \system{} consistently achieves high attack success rates. Additionally, \system{} reduces queries by up to 80\% in black-box scenarios and extends to cases where CLIP is unfamiliar with specific textual concepts.
\end{itemize}


\section{Background and Related Work}

\subsection{Vision-Language Models (VLMs)}

VLMs, such as CLIP \cite{openai2021clip}, are notable for learning visual and textual representations through contrastive pre-training on large-scale image-text pairs. In CLIP, a text encoder \(E_T(x)\) and an image encoder \(E_I(x)\) are pre-trained to match texts with corresponding images. CLIP can be applied to tasks such as zero-shot classification, which is defined as follows:
\begin{equation}
F(x) = \text{argmax}_{y_k \in \{y_1, y_2, ..., y_N\}} \text{sim}(E_I(x), E_T(y_k))
\end{equation}
where \(x\) is the image input, \( \{y_1, y_2, ..., y_N\} \) are text labels for different classes, and \(\text{sim}(\cdot,\cdot)\) denotes cosine similarity. This equation compares the embeddings of the input image and different labels, selecting the label with the highest similarity as the predicted label. Since CLIP is pre-trained on billions of image-text pairs, it can perform zero-shot classification on many tasks with high accuracy.

In the security domain, VLM is also used in many works for attacks. However, previous works either attack the VLM itself \cite{fort2021pixels, luo2024an, zhang2024universal} or its downstream applications \cite{bagdasaryan2024adversarial}. In contrast, our threat model diverges significantly, aiming to attack various models for specific tasks using textual concepts and VLMs, even when these victim models are unrelated to VLMs.

\subsection{Targeted Adversarial Attacks}

Targeted adversarial attacks in machine learning involve crafting inputs to a model designed to cause it to predict a specific target class. These attacks exploit vulnerabilities in the model's processing of input data, leading to incorrect outputs while the inputs may appear unchanged to human observers. A typical targeted adversarial attack can be formulated as:
\begin{equation}
x^* = \arg \min_{x^*} (\ell(f(x^*), y_t)), \text{ s.t. } \|x - x^*\| \leq \epsilon,
\end{equation}
where \(x^*\) is the adversarial example, \(\ell\) is the loss function, and \(\epsilon\) is the perturbation strength constraint. \(y_t\) is the target class. Various forms of adversarial attacks are relevant to our work.

\subsubsection{White-box Adversarial Attacks}

White-box adversarial attacks assume complete knowledge of the target model, including its architecture, parameters, and gradients. This comprehensive access allows attackers to craft perturbations by directly optimizing an objective function to maximize misclassification. Classic methods such as the Fast Gradient Sign Method (FGSM) \cite{goodfellow2014explaining} and Projected Gradient Descent (PGD) \cite{madry2017towards} utilize gradient information to create perturbations that induce high attack efficacy with minimal distortion. Advanced approaches like the C\&W attacks \cite{carlini2017towards} optimize more sophisticated loss functions to bypass defensive mechanisms. \partrevisionH{Agrawal et al. \cite{agrawal2023m} propose M-SAN, a patch-based transferable attack under black-box settings, using a multi-stack adversarial network to target face recognition models effectively.} White-box attacks serve as a benchmark for evaluating model robustness, representing an upper bound on adversarial vulnerability due to their extensive access to model information.

\subsubsection{\partrevisionH{Decision-based} Black-box Adversarial Attacks}

\partrevisionH{Decision-based} black-box adversarial attacks operate without access to the model's architecture, parameters, or training data. Instead, they typically rely on query feedback to craft perturbations. Liu et al. \cite{liu2016delving} introduced a method to generate adversarial examples using a substitute model trained to mimic the target's decision boundaries. Chen et al. \cite{chen2020signflip} proposed a query-efficient black-box attack using sign flips to accelerate the search. Despite such advances, state-of-the-art methods often demand over 10,000 queries to attack a specific class when only discrete label predictions are available \cite{zheng2023blackboxbench}. This work investigates textual concepts as a more practical and efficient approach to adversarial attacks. \partrevisionH{Some recent works \cite{liu2023boosting} use gradient priors, which seamlessly integrate the data-dependent gradient prior and time-dependent prior into the gradient estimation procedure, reducing the required queries to 5,000 to achieve a 36\%–48\% ASR on ImageNet.}

\subsubsection{Transferable Adversarial Attacks}

Unlike white-box and black-box adversarial attacks, transferable adversarial attacks fall under the category of ``no-box'' attacks \cite{li2020practical, chen2017zoo} as they do not require access to the target model in any capacity. Transferable adversarial attacks typically assume that adversaries control a surrogate model that closely resembles the target model. These attacks can be broadly categorized into two types: sample-specific transferable attacks and sample-agnostic transferable attacks.

\noindent \textbf{Sample-Specific Transferable Attacks.} These attacks focus on optimizing perturbations for each input sample to maximize transferability across unseen models. Notable works in this area include AA \cite{inkawhich2019feature}, which aligns high-level feature representations of source and target images to craft highly transferable targeted examples. Logit \cite{zhao2021success} and Logit Margin \cite{weng2023logit} introduce loss functions that use logit- or margin-based objectives to address vanishing gradient issues in targeted attacks, thereby enhancing transferability. SU \cite{wei2023selfuniversal} demonstrates the advantage of creating “self-universal” perturbations across regions within a single image, boosting transfer performance without additional training data. FFT \cite{zeng2024enhancing} employs feature-space fine-tuning, starting from a baseline adversarial example, to emphasize features of the target class while suppressing those of the original class, further improving transferability.

\noindent \textbf{Sample-Agnostic Transferable Attacks.} These attacks aim to generate \textit{universal adversarial perturbations} or train \textit{generative models} applicable to multiple inputs, thereby avoiding individual optimization. TTP \cite{naseer2021ttp} trains a generator to align perturbed image distributions with a target class, achieving highly transferable targeted attacks without reliance on class boundaries. C-GNC \cite{fang2024clip} employs a class-conditional generator for multi-target attacks, integrating semantic cues from CLIP through cross-attention mechanisms to significantly boost success rates. Universal approaches like CleanSheet \cite{ge2024hijacking} create a single perturbation applicable to any input by using robust features from the target model’s clean training data to induce misleaded predictions. In the neighboring backdoor setting, generative trigger optimization has also been used to identify naturally occurring image features that improve both stealth and attack potency \cite{xu2026breaking}. That training-time poisoning setting is distinct from ours, which does not modify the victim's training set, but it further illustrates how optimizing reusable trigger patterns can strengthen an attack across inputs.

\begin{figure*}
\begin{center}
\centerline{\includegraphics[width=0.9\textwidth]{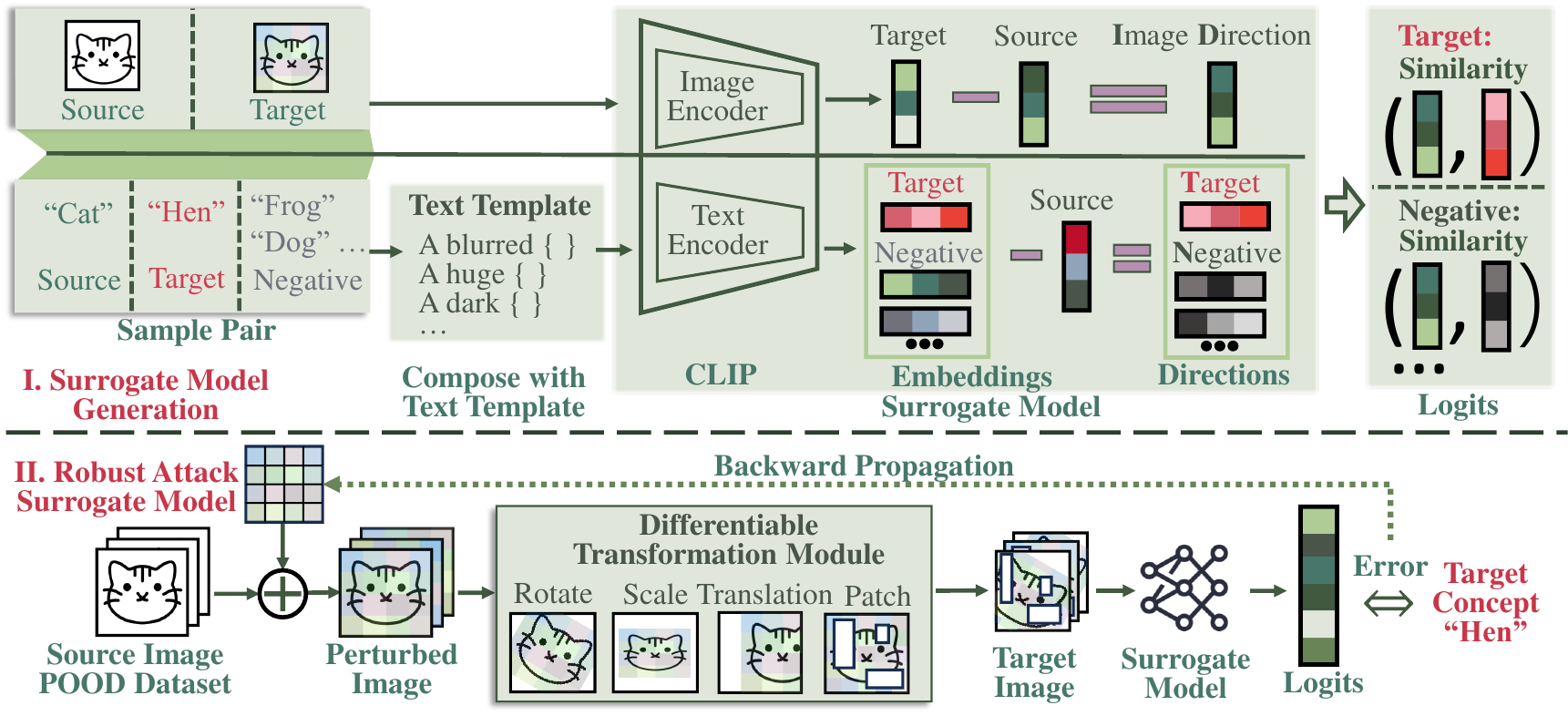}}
\vspace{-0.2cm}
\caption{Overview of \system{}.}
\vspace{-0.4cm}
\label{fig:overview}
\end{center}
\end{figure*}

\subsubsection{Vision-Language Model Attacks}

With the rapid development of vision-language models, recent works have explored transferable adversarial attacks targeting these models. Several studies \cite{bagdasaryan2023ab, carlini2024aligned, qi2024visual} utilize adversarial perturbations for jailbreaking and prompt injection in multimodal chatbots, which are specific downstream tasks. In \cite{shayegani2023plug, wang2023instructta}, attackers are assumed to control a vision encoder, allowing attacks to affect several downstream tasks that utilize this encoder, which can be regarded as grey-box transferable attacks. Dong et al. \cite{dong2023robust} present an \textit{untargeted} attack on image embeddings that causes the model to incorrectly predict the main object in an image. Additionally, \cite{bagdasaryan2024adversarial} introduces a multimodal embedding adversarial attack method to misalign inputs with arbitrary target inputs from different modalities. These targeted attacks can compromise all downstream tasks that use the same or similar embedding models. Our approach differs significantly from these existing methods. Instead of targeting specific models or specific downstream tasks, we aim to attack a diverse range of vision models, including those used for image classification, image searching, image generation, and vision-language tasks, without making assumptions about their architectures or specific applications.

\section{Threat Model}

\label{sec:threat_model}

\subsubsection{\textbf{Adversary’s Goals.}} The adversary aims to create a universal perturbation $\mathcal{T}$, constrained by an $l_{\infty}$-norm $\epsilon$, that causes as many images as possible from the target test distribution $X_t$ to be classified by the target model $f$ as a specific target class $y_t$. The optimization goal minimizes the expected targeted loss over $X_t$ and does not require success on every individual input:
\begin{equation}
\mathcal{T}^{*} = \arg\min \mathbb{E}_{x \sim X_t} \ell(f(x+\mathcal{T}), y_{t}), \quad \text{s.t.} \ \| \mathcal{T} \|_{\infty} \leq \epsilon,
\label{eq:formulation}
\end{equation}
where $\ell$ represents the loss function. This objective follows the standard formulation of universal adversarial examples.

\subsubsection{\textbf{Adversary’s Knowledge.}} (1) The attacker is assumed to know the complete set of labels \( Y \), defined by specific words or phrases. These labels consist of target concepts (target class) and negative concepts (non-target classes). The attacker also has a basic understanding of the victim's learning task, including operational resolution. (2) The attacker is presumed to have access to a public dataset relevant to the task, which helps in crafting the attack. However, this dataset does not need to share the same distribution or labels as the victim’s training data. In some cases, the public and the target datasets may differ significantly in labels, preprocessing, and normalization. We refer to such public datasets as \textit{public out-of-distribution (POOD) datasets}. (3) Moreover, the attacker is believed to have access to a public general CLIP \cite{openai2021clip} model. CLIP is known to connect images and texts within the same domain and encapsulate the knowledge required for the victim’s task.

\section{Methodology}

\subsection{Overview}

Our goal is to craft a perturbation capable of adversarially attacking the target model using its textual labels. This task can be formulated as an optimization problem in Eq. \ref{eq:formulation}. However, two critical components are assumed inaccessible: the target model \(f\) and the target dataset \(X_t\). To overcome these challenges: \textbf{(a)} we address the missing target model by aligning the public general model with the target model using textual concepts, termed Model Alignment. \textbf{(b)} We tackle potential inner biases in the public dataset by using a public dataset and novel embedding reduction techniques, termed Dataset Alignment. Finally, \textbf{(c)} to ensure the perturbation is transferable and robust, we apply strong random transformations to each image fed into the surrogate model.

In conclusion, the pipeline of \system{}, illustrated in Fig. \ref{fig:overview}, can be divided into two parts: \textbf{(I)} building an aligned surrogate model with CLIP to simulate the behavior of the target model while mitigating inner biases in the input data, and \textbf{(II)} performing a targeted transferable adversarial attack on this surrogate model to obtain a transferable perturbation.

\begin{figure}[t]
\vspace{-0.2cm}
\begin{small}
\begin{algorithm}[H]
\caption{Generate Perturbation with \system{}.}
\label{alg:perturbation}
\begin{algorithmic}[1]
\State \textbf{Input:} public dataset \(x, y \sim \mathcal{X}, \mathcal{Y}\), max step \(\mathcal{N}\), target concept \(y_t\), CLIP image encoder \(E_I\), CLIP text encoder \(E_T\), negative concepts \(Y_n\), \(l_{\infty}\) constraint \(\epsilon\), learning rate \(\alpha\)
\State \textbf{Output:} Perturbation \(\mathcal{T}\)

\Function{surrogate\_model}{$\hat{x}, x, y, y_t, Y_n$}
    \State \(\vec{x} = E_I(\hat{x}) - E_I(x)\)
    \State \(\vec{y}_t = E_T(y_t) - E_T(y)\)
    \State \(\vec{Y}_n = E_T(Y_n) - E_T(y)\)
    \State \textbf{return} \(\text{sim}(\vec{x}, \vec{y}_t), \text{sim}(\vec{x}, \vec{Y}_n)\)
\EndFunction

\State Initialize \(\mathcal{T} \sim \text{Gaussian}(0, 1)\);
\For{\(n = 1\) to \(\mathcal{N}\)}
    \State Sample a batch of clean examples \(x, y\) from \(\mathcal{X}, \mathcal{Y}\)
    \State \(\hat{x} = DT(x + \mathcal{T})\) \Comment{Differentiable Transformation}
    \State \( \text{logits} = \Call{surrogate\_model}{\hat{x}, x, y, y_t, Y_n} \)
    \State \(\mathcal{L} = - \text{log-likelihood}(\text{logits})\) \Comment{Using Eq. \ref{eq:loss}}
    \State \(\mathcal{T} \leftarrow \mathcal{T} - \alpha \cdot \nabla_{\mathcal{T}}\mathcal{L}\)
    \State \(\mathcal{T} \leftarrow \text{clamp}(\mathcal{T}, \epsilon)\)
\EndFor
\State \textbf{return} \(\mathcal{T}\)
\end{algorithmic}
\end{algorithm}
\end{small}
\vspace{-0.5cm}
\end{figure}

Algorithm \ref{alg:perturbation} details the workflow of \system{}. \partrevisionH{The algorithm takes as input a public dataset, the target concept, the CLIP encoders, negative concepts, and hyperparameters such as the maximum number of steps, the \(l_{\infty}\) constraint, and the learning rate, producing an optimized perturbation \(\mathcal{T}\). The process is as follows: \textbf{Initialization (Line 3)}: The perturbation \(\mathcal{T}\) is initialized with small random values from a Gaussian distribution. \textbf{Optimization Loop (Lines 4-12)}: For each iteration up to \(\mathcal{N}\), a batch of clean examples \(x, y\) is sampled from the public dataset, the perturbation is added to these images, and the resulting perturbed images are processed through a differentiable transformation module (Line 12) to enhance robustness. The surrogate model (Lines 4-6) computes similarities between the image direction (perturbed minus clean image embeddings) and text directions (target/negative concept embeddings minus source concept embedding). The loss is calculated using the negative log-likelihood of the target similarity relative to negative similarities (Line 13), and the perturbation is updated via gradient descent (Line 14) and clamped to the \(l_{\infty}\) bound (Line 15). This iterative process refines the perturbation to maximize the likelihood of images being classified as the target concept while ensuring robustness to transformations.} Lines 4-6 show how feature direction helps mitigate potential biases. Lines 3-8 detail the construction of a surrogate model based on direction similarity. Line 12 describes the differentiable transformation applied to the input with perturbation.

\subsection{Build surrogate model with text concepts}

\subsubsection{\textbf{Input.}} The constructed surrogate model takes multiple inputs, as shown in Fig. \ref{fig:overview}: the source image \(x\), the perturbed image \(\hat{x}\), the source concept \(y\), the target concept \(y_t\), and the negative concepts \(Y_n\). The target concept corresponds to the target class in text format, while the negative labels represent non-target classes in text format. Note that we do not assume any source concept \(y\) exists in negative concepts \(Y_n\) or target concept $y_t$.

\subsubsection{\textbf{Text Template Composition.}} \partrevisionC{To ensure that textual concepts are effectively encoded and aligned with CLIP's training data, we process all source, target, and negative concepts through a random text template module. This module applies varied templates, such as ``a photo of a [concept]'', ``a blurry image of a [concept]'', or ``a pixelated version of a [concept]'', mimicking the diverse text descriptions seen during CLIP’s training. This alignment ensures that the text embeddings accurately represent the concepts in a format familiar to CLIP. By applying multiple templates and averaging their embeddings, we enhance the robustness of these representations, making them less sensitive to specific phrasings and more generalizable across contexts. This step is essential for obtaining reliable embeddings, which are critical for building the surrogate model and generating the adversarial perturbation.}


\subsubsection{\textbf{Bias Alignment in Dataset.}} 
\label{sec:bias_align}
After preprocessing the textual concepts, both image and text are passed to the CLIP image encoder \(E_I(\cdot)\) and text encoder \(E_T(\cdot)\) to obtain embeddings for the image and text. The embedding is a 1-D vector that represents the latent information of a single image or an entire text sentence. In CLIP, both image and text embeddings are trained in a contrastive manner within the same latent space, ensuring that different modalities are aligned. We then apply the concept of direction to calculate the difference between the target embedding and the source embedding vectors. This forms the image direction \(D_x\) and text directions \(D_y\) (which includes both the target text direction \(D_{y_t}\) and negative text directions \(D_{Y_n}\)), calculated as follows:
\begin{align}
\vspace{-0.05cm}
\label{eq:direction_1}
\quad D_x &= E_I(\hat{x}) - E_I(x), \\
\label{eq:direction_2}
\quad D_{y_t} &= E_T(y_t) - E_T(y), \\
\label{eq:direction_3}
\quad D_{Y_n} &= E_T(Y_n) - E_T(y)
\end{align}

A key design in this part is to use the directions of the target and negative embeddings instead of the embeddings themselves. This approach helps eliminate inherent biases within the input dataset. For example, if a bias \(\sigma\) exists in an image (e.g., the entire image has a slight blue tint), the image embedding becomes \(E_I(\hat{x} + \sigma)\). If we optimize directly using this embedding, \(\hat{x}\) will attempt to offset this bias to be \(\hat{x} = \hat{x}_\text{correct} - \sigma\), thereby reducing its applicability to unbiased inputs. In contrast, using the image direction results in \(E_I(\hat{x} + \sigma) - E_I(x + \sigma)\), which significantly mitigates the impact of bias, as the biases are subtracted. \partrevisionF{This is because the bias \(\sigma\) is present in both \(E_I(\hat{x} + \sigma)\) and \(E_I(x + \sigma)\), so their difference \(D_x = E_I(\hat{x} + \sigma) - E_I(x + \sigma) \approx E_I(\hat{x}) - E_I(x)\), assuming that the bias affects both embeddings additively. Thus, \(D_x\) primarily reflects the change due to the perturbation \(T\), independent of \(\sigma\).} This approach makes the perturbation more generalizable and input-agnostic.

\subsubsection{\textbf{Logit Calculation.}} After getting the image direction and text directions, we calculate the cosine similarity \(\text{sim}(\cdot, \cdot)\) between directions: 
\begin{align}
\vspace{-0.05cm}
\label{eq:sim_1}
\quad \text{sim}(D_x, D_{y_t}) &= \frac{D_x \cdot D_{y_t}}{\|D_x\| \|D_{y_t}\|}, \\
\label{eq:sim_2}
\quad \text{sim}(D_x, D_{y_N}) &= \frac{D_x \cdot D_{y_N}}{\|D_x\| \|D_{y_N}\|}.
\end{align}
\partrevisionC{These similarities serve as logits for the target and negative concepts, where \(\text{sim}(D_x, D_{y_t})\) measures alignment with the target concept, and \(\text{sim}(D_x, D_{Y_n})\) measures alignment with negative concepts (all non-target labels, including the true label). By maximizing \(\text{sim}(D_x, D_{y_t})\) and minimizing \(\text{sim}(D_x, D_{Y_n})\), the perturbation increases similarity to the target concept while decreasing similarity to non-target concepts, aligning with CLIP’s contrastive learning objective of distinguishing matching from non-matching pairs. This general approach enhances transferability by ensuring the perturbation generalizes across concepts and models, rather than overfitting to the true label or specific dataset features.}


\subsection{Robust Attack on Surrogate Model}

\subsubsection{\textbf{Universal Adversarial Perturbation (UAP).}} We perform adversarial attacks using a Universal Adversarial Perturbation (UAP) approach \cite{hirano2020tuap}, where the perturbation \(\mathcal{T}\) remains consistent across different images and architectures. The perturbed image in UAP is defined as \(\hat{x} = x + \mathcal{T}\). The perturbation \(\mathcal{T}\) is optimizable and incorporates weight decay to prevent overfitting.

\subsubsection{\textbf{Differentiable Transformation Module.}} Perturbed images are processed through a differentiable transformation module to enhance robustness and transferability, as inspired by prior work \cite{wei2023selfuniversal}. Specifically, we adapt five transformations: rotation, scaling, translation, patching, and horizontal flipping. Each transformation is applied randomly with varying strength and parameters, resulting in a target image that is a composite of all these transformations and significantly different from the original image. Notably, random patching is a unique design in our module. We create three non-overlapping rectangular patches with random positions, widths, and heights to obscure the original information. This approach helps avoid local dependencies and encourages the perturbation to capture more global patterns. Additionally, if any pixel information is lost during transformations, such as patching or extending beyond the image boundary, we fill those areas with zeros.

\subsubsection{\textbf{Loss Function.}} When target images are passed to the surrogate model and yield logits, we normalize these raw logits to probabilities using the Softmax function. Since our objective is to maximize the likelihood of the target image being predicted as the target concept, we employ the Negative Log-Likelihood (NLL) as the loss function, calculated as follows:
\begin{equation} 
\label{eq:loss}
\mathcal{L} = -\log \left( \frac{e^{\text{sim}(D_x, D_{y_t})}}{\sum_{i=1}^{M} e^{\text{sim}(D_x, D_{y_N})^{(i)}} + e^{\text{sim}(D_x, D_{y_t})}} \right)
\end{equation}
Here, \(\text{sim}(D_x, D_{y_t})\) and \(\text{sim}(D_x, D_{y_N})\) represent the similarities of the target and negative concepts, respectively, as detailed in the previous section. \(M\) denotes the total number of negative concepts. This approach allows us to maximize the likelihood of the perturbed image being predicted as the target class in a robust manner.

\section{Evaluation}

\begin{figure*}
\begin{center}
\centerline{\includegraphics[width=0.9\textwidth]{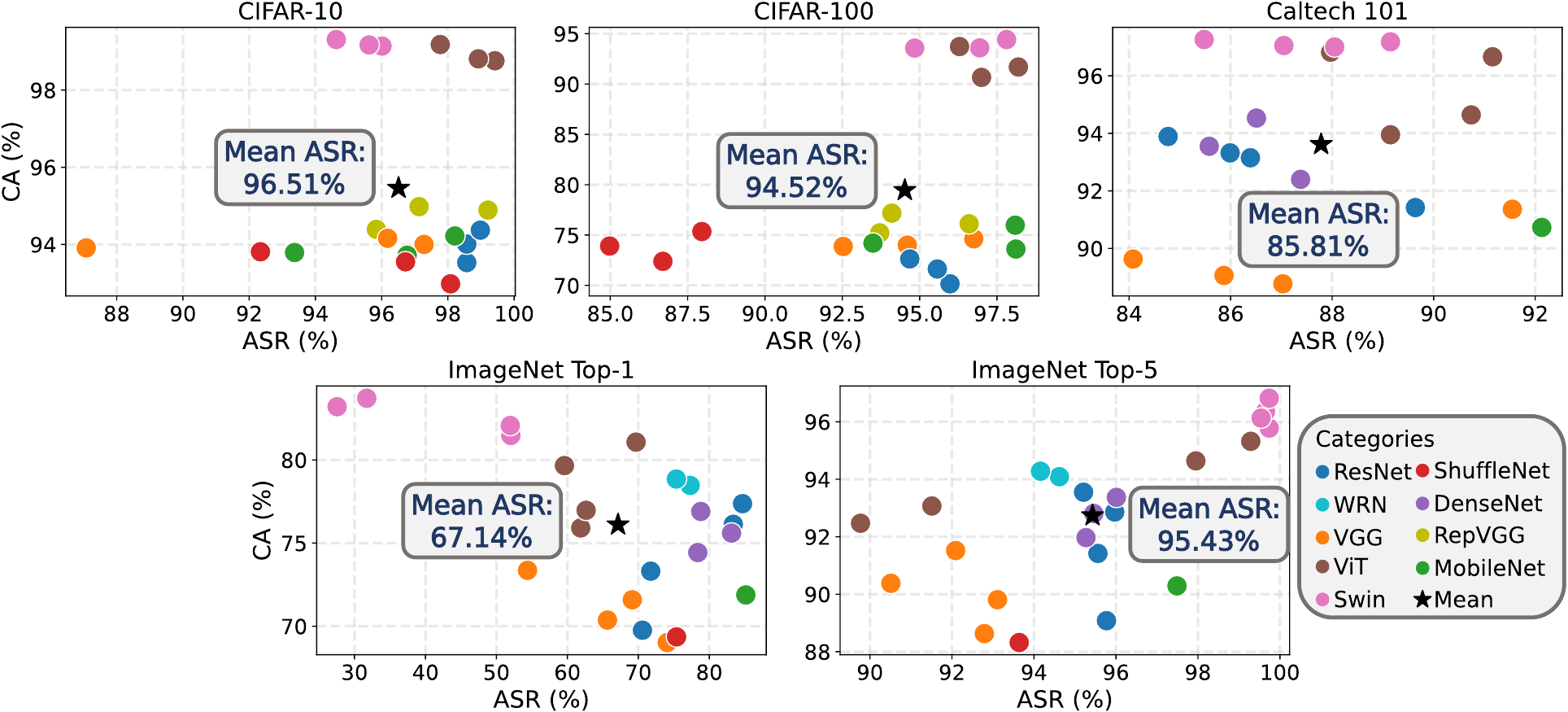}}
\vspace{-0.3cm}
\caption{Attack Performance. \system{} can transfer attacks to all models across all datasets using only CLIP.}
\vspace{-0.4cm}
\label{fig:base_all_attack}
\end{center}
\end{figure*}

\subsection{Experiment Setup}

\textbf{Datasets.} Our experiments involve four image datasets: CIFAR-10 \cite{krizhevsky2009cifar10}, CIFAR-100 \cite{krizhevsky2009cifar10}, Caltech-101 \cite{fei2006caltech}, and ImageNet \cite{le2015imagenet}. Since our attack does not access the target dataset, our perturbation is optimized using a public out-of-distribution (POOD) dataset. We use Tiny-ImageNet \cite{le2015tiny} as the POOD dataset for both CIFAR-10 and CIFAR-100. For Caltech-101 and ImageNet, we use ImageNet and ImageNet-21K, respectively. Specifically, we remove overlapping classes in ImageNet-21K to exclude ImageNet-1K classes, ensuring no class or image overlap between the target training set and the POOD set. Details of datasets and target classes for each training pipeline are provided in Table \ref{tab:dataset_setting}.

\begin{table}
\centering
\caption{Experimental setup for all datasets. Perturbations are optimized on the public out-of-distribution (POOD) dataset and tested on the target dataset. ImageNet-21K is modified to exclude labels from ImageNet-1K to avoid label overlap.}
\vspace{-0.3cm}

\label{tab:dataset_setting}
\begin{small}
\begin{tblr}{
  width = \linewidth,
  rowsep = 0.8pt,
  colsep = 0.4pt,
  colspec = {Q[196]Q[190]Q[180]Q[165]Q[180]},
  cells = {c},
  hline{1,7} = {-}{0.1em},
  hline{2} = {1}{lr},
  hline{2} = {2}{l},
  hline{2} = {3-4}{},
  hline{2} = {5}{r},
}
{\textbf{Dataset}} & {\textbf{CIFAR-10}} & {\textbf{CIFAR-100}} & {\textbf{Caltech-101}} & {\textbf{ImageNet}} \\
{\textbf{\# of Classes}}    & {10}                & {100}                & {101}                  & {1000}              \\
{\textbf{Input Shape}}      & {(3, 32, 32)}       & {(3, 32, 32)}        & {(3, 224, 224)}        & {(3, 224, 224)}     \\
{\textbf{Total Images}}     & {60,000}            & {60,000}             & {9,146}                & {1,431,167}         \\
{\textbf{POOD Dataset}}     & {TinyImageNet}     & {TinyImageNet}      & {ImageNet}             & {ImageNet-21K}        \\
{\textbf{Target Class}}     & {8 (Ship)}          & {8 (Bicycle)}        & {8 (Barrel)}           & {8 (Hen)}        
\end{tblr}
\end{small}
\vspace{-0.3cm}
\end{table}

\noindent \textbf{Metrics.} In our experiments, we use two metrics: \textit{Clean Accuracy} (CA) and \textit{Attack Success Rate} (ASR). CA measures the model's classification accuracy on unperturbed data, while ASR indicates the percentage of test instances with embedded perturbations that are classified as the target class by the model.

\noindent \textbf{Models.} We use OpenCLIP's implementation of CLIP \cite{openai2021clip}, configured with ViT-B-32 and pretrained on the Laion2B dataset \cite{schuhmann2022laion}, as the default surrogate model across all datasets. The differentiable transformation module applies ±5° rotations, ±5\% translations, scaling (0.95x–1.05x), horizontal flips with a probability of 50\%, and randomly patches three rectangular regions per image. For universal perturbation optimization, we employ the Adam optimizer \cite{kingma2014adam}, setting the learning rate to 0.01, weight decay to \(1 \times 10^{-5}\), batch size to 64, and training for 5000 steps. We set the $l_{\infty}$-norm of perturbations to be upper bound by 32/255 as a default setting, which is a standard setting in both transferable adversarial perturbation \cite{naseer2019crossdomain, li2020regional, naseer2021generating} and backdoor learning \cite{zeng2023narcissus}.

A total of 85 pre-trained models are used as the targets to evaluate our attack, including 27 models for CIFAR-10, 27 for CIFAR-100, 24 for Caltech-101, and 23 for ImageNet. All the models for ImageNet are obtained from Torchvision without any modifications. All the models' weights for CIFAR-10 and CIFAR-100 except ViT and Swin are directly loaded from GitHub. For ViT and Swin, we finetuned for 7 epochs with pre-trained weight on HuggingFace. For Caltech-101, we train 14 models under normal settings, respectively.

\subsection{Attack Performance.}

\textbf{Attack Success Rate.} The ASRs are reported across several datasets, including CIFAR-10, CIFAR-100, Caltech-101, and ImageNet. These results demonstrate the effectiveness of the proposed method in attacking multiple neural networks using textual concepts. As shown in Fig. \ref{fig:base_all_attack}, low-resolution datasets like CIFAR-10 and CIFAR-100 have high ASRs, averaging 96.51\% and 94.52\%, respectively. In high-resolution datasets, the average ASR exceeds 85\%, reaching up to 92.13\% in Caltech-101. In contrast, ImageNet has a top-1 accuracy of 67.14\%. While this is lower than other datasets, it highlights the challenges posed by ImageNet's 1000 classes, where many labels are semantically similar (e.g., ``hen'' vs. ``rooster'', ``partridge'', ``limpkin''). This complexity makes top-1 accuracy less indicative of overall attack success. As a result, we also consider top-5 accuracy in ImageNet, which shows an average ASR of 95.43\%, indicating the attack's ability to mislead models across multiple classifications.

\noindent \textbf{Transferability.} The transferability of \system{} is validated by its effectiveness across different model structures. This indicates that adversarial perturbations can work well beyond the surrogate model. As shown in Fig. \ref{fig:base_all_attack}, the lowest mean ASRs across all four tested datasets remain impressively high at 67\%. Notably, even advanced models like ViT and Swin Transformer—typically challenging for transferable attacks \cite{gusurvey}—still achieve excellent performance with our approach, highlighting its strong transferability.

\begin{table}
\centering
\caption{ASR of different target classes on CIFAR-10. Our attack is robust to different target classes.}
\vspace{-0.3cm}
\label{tab:target_cifar10}

\begin{small}

\begin{tblr}{
  width = \linewidth,
  rowsep = 0.8pt,
  colsep = 0.8pt,
  colspec = {Q[300]Q[127]Q[127]Q[127]Q[127]Q[127]},
  cells = {c},
  hline{1,10} = {-}{0.10em},
  hline{2,9} = {-}{0.05em},
}
\textbf{Target Class→} & \textbf{1 (car)} & \textbf{3 (cat)} & \textbf{5 (dog)} & \textbf{7 (horse)} & \textbf{9 (truck)} \\
ResNet                & 94.99      & 94.88      & 93.41      & 90.89      & 99.62      \\
VGG                   & 94.80      & 94.49      & 84.08      & 80.85      & 97.45      \\
MobileNet             & 94.53      & 97.42      & 85.75      & 90.24      & 99.83      \\
ShuffleNet            & 92.59      & 92.92      & 80.70      & 90.40      & 99.09      \\
RepVGG                & 96.71      & 98.37      & 93.28      & 89.51      & 99.48      \\
ViT                   & 98.88      & 98.52      & 98.16      & 96.09      & 97.74      \\
Swin                  & 96.91      & 97.70      & 94.94      & 95.91      & 95.96      \\
Average                 & 95.63      & 96.33      & 90.05      & 90.56      & 98.45      
\end{tblr}
\end{small}
\vspace{-0.2cm}
\end{table}

\noindent \textbf{Robustness to target class.} To assess the robustness of our method to various targets, we conducted several attacks targeting different classes. As shown in Table \ref{tab:target_cifar10}, we present the average performance across structures with different scales (e.g., the ASR for ResNet represents the average ASR for ResNet-20, ResNet-32, ResNet-44, and ResNet-56). All target classes achieved an average ASR of over 90\%, demonstrating strong robustness in target selection.

\begin{figure}
\begin{center}
\centerline{\includegraphics[width=1.0\columnwidth]{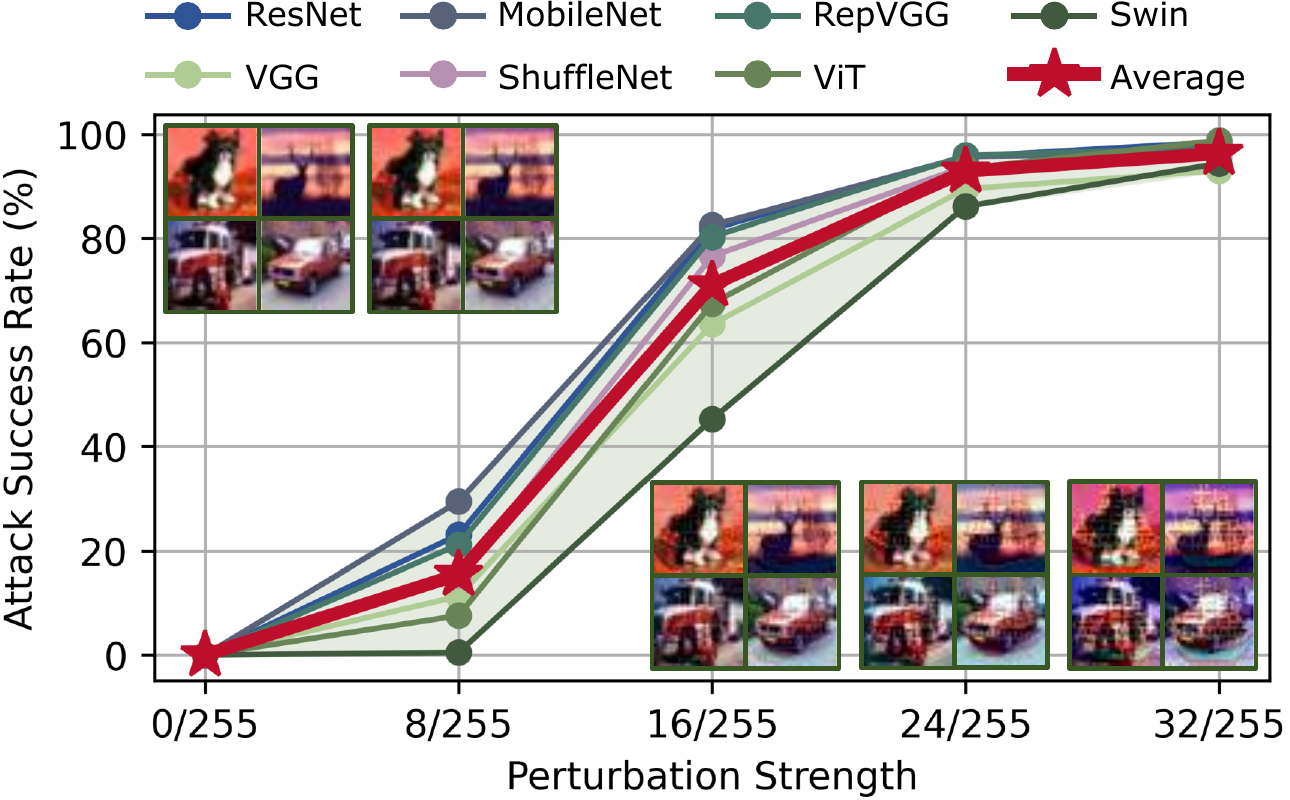}}
\vspace{-0.35cm}
\caption{Perturbation's constraints and ASR on CIFAR-10.}
\label{fig:norm}
\vspace{-0.4cm}
\end{center}
\end{figure}

\begin{table*}[t]
\centering
\begin{small}
\caption{Comparison of ASR with other attacks on ImageNet. Each table cell shows \textit{top-1 ASR / top-5 ASR}, with top-1 ASR above 30\% highlighted in red. ``RN50(IM)'' denotes a ResNet-50 model pretrained on ImageNet, while ``CLIP'' refers to a zero-shot CLIP classifier. Although all baselines perform well using RN50 as the surrogate model, they fail on CLIP. In contrast, our method is the only one that remains effective and achieves strong transferability when using CLIP.}
\vspace{-0.3cm}
\label{tab:comparison}

\begin{tblr}{
  width = 0.9\linewidth,
  colspec = {Q[85]Q[83]Q[83]Q[83]Q[83]Q[83]Q[83]Q[83]Q[83]Q[110]Q[83]},
  rowsep = 0.8pt,
  colsep = 0.8pt,
  cells = {c},
  cell{3}{2-4,6,8-11} = {bg = customred},
  cell{3}{5,7,9}      = {bg = customgreen},
  cell{4}{3-4,6-11}   = {bg = customred},
  cell{4}{2,5,7}      = {bg = customgreen},
  cell{5}{2-4,6,8-11} = {bg = customred},
  cell{5}{5,7,9}      = {bg = customgreen},
  cell{6}{6,8-11}     = {bg = customred},
  cell{6}{2-5,7,9}    = {bg = customgreen},
  cell{7}{2-4,6-11}   = {bg = customred},
  cell{7}{5,7}        = {bg = customgreen},
  cell{8}{9,11}       = {bg = customred},
  cell{8}{2-8,10}     = {bg = customgreen},
  cell{9}{6,8,11}     = {bg = customred},
  cell{9}{2-5,7,9-10}= {bg = customgreen},
  cell{10}{3-4,6,8-11}= {bg = customred},
  cell{10}{2,5,7,9}   = {bg = customgreen},
  cell{1}{1} = {r=2}{},
  cell{1}{4} = {c=2}{},
  cell{1}{6} = {c=2}{},
  cell{1}{8} = {c=2}{},
  hline{1,11} = {-}{0.1em},
  hline{2} = {2-11}{0.05em},
  hline{10} = {-}{0.05em},
  hline{3} = {-}{},
  hline{2} = {2,3,4,6,8,10,11}{l},
  hline{2} = {2,3,5,7,9,10}{r},
}
{Surrogate\\Model→} & Logit \cite{zhao2021success} & SU \cite{wei2023selfuniversal}   & Logit Margin \cite{weng2023logit} &           & FFT \cite{zeng2024enhancing}  &           & CGNC \cite{fang2024clip} &           & CleanSheet \cite{ge2024hijacking} & Ours      \\
                    & RN50(IM)   & RN50(IM)  & RN50(IM)          & CLIP      & RN50(IM)  & CLIP      & RN50(IM)  & CLIP      & Ensemble        & CLIP      \\
ResNet              & 46.1/67.0  & 71.2/85.1 & 58.2/78.8         & 7.6/21.0  & 82.2/95.9 & 9.1/25.1  & 80.9/93.2 & 22.1/44.1 & 63.0/79.0       & 77.2/95.2 \\
VGG                 & 27.0/53.2  & 38.5/60.2 & 35.6/62.8         & 1.2/4.1   & 78.6/95.7 & 1.6/7.3   & 80.8/95.4 & 33.9/55.2 & 66.8/79.8       & 65.8/92.1 \\
MobileNet           & 35.0/60.8  & 51.7/72.3 & 44.3/70.4         & 7.9/22.1  & 83.3/95.7 & 12.2/31.9 & 85.0/92.7 & 14.3/34.9 & 65.2/72.4       & 85.1/97.5 \\
ShuffleNet          & 22.9/47.7  & 15.4/35.3 & 19.2/37.3         & 6.2/14.2  & 58.4/82.4 & 9.3/22.6  & 56.9/78.6 & 10.5/24.2 & 52.4/87.6       & 75.4/93.6 \\
DenseNet            & 54.3/79.7  & 81.0/92.8 & 79.3/92.8         & 8.0/23.5  & 94.5/99.0 & 11.0/27.6 & 88.4/96.6 & 45.3/59.9 & 70.4/70.5       & 80.1/95.6 \\
ViT                 & 6.8/24.6   & 7.6/24.1  & 24.5/51.0         & 23.5/41.6 & 26.5/53.3 & 23.6/41.4 & 27.5/52.1 & 60.7/77.6 & 24.7/52.0       & 63.4/94.6 \\
Swin                & 4.9/21.1   & 12.5/41.5 & 23.4/44.9         & 11.2/24.3 & 33.9/61.2 & 11.7/26.4 & 49.1/78.9 & 10.7/50.9 & 27.6/55.5       & 40.8/99.7 \\
Average             & 28.2/50.6  & 39.7/58.8 & 40.6/62.6         & 9.4/21.5  & 65.3/83.3 & 11.2/26.0 & 66.9/83.9 & 28.2/49.5 & 52.9/71.0       & 69.7/95.5 \\
\end{tblr}
\vspace{-0.2cm}
\end{small}
\end{table*}

\subsection{Perturbation's Visibility.}

\noindent \textbf{ASR vs. Perturbation Strength.} In this paper, we use the \( l_{\infty} \)-norm to constrain our perturbation. To explore the impact of different constraints on performance, we conducted experiments with constraints of \( \frac{8}{255} \), \( \frac{16}{255} \), \( \frac{24}{255} \), and \( \frac{32}{255} \). The results, including the ASR and sample images demonstrating perturbation visibility, are shown in Fig. \ref{fig:norm}. The ASR only decreases slightly, from 96.3\% to 92.9\%, when the \( l_{\infty} \)-norm is reduced from \( \frac{32}{255} \) to \( \frac{24}{255} \). The ASR remains high at 71.1\% even when the \( l_{\infty} \)-norm drops to \( \frac{16}{255} \).

\begin{partGsection}
\noindent \textbf{Stealthiness vs. Perturbation Strength.} 
Following prior work \cite{ge2024hijacking}, we conducted a human study to assess the stealthiness of \system{}. Our evaluation focused on determining whether adversarial inputs generated by \system{} remain stealthy. To this end, we designed a survey involving 10 volunteers with expertise in adversarial robustness. For the experiment, we generated 100 adversarial inputs with perturbation strengths ranging from \( \frac{8}{255} \) to \( \frac{32}{255} \). Each sample was perturbed to target a randomly selected class using \system{}. Participants were tasked with identifying the category of each sample and classifying the input as "normal," "abnormal," or "without visible triggers."

As illustrated in Fig. \ref{fig:human}, the results reveal that as perturbation strength decreases, volunteers increasingly perceive the inputs as normal and invisible perturbations. At a perturbation strength of \( \frac{16}{255} \), nearly all participants rated the perturbed images as normal, with 64.5\% of samples deemed to have invisible perturbations. Conversely, at a higher perturbation strength of \( \frac{32}{255} \), the perturbations become largely perceptible, though only 32.7\% of samples were classified as abnormal. Furthermore, recognition accuracy remained consistently high across all perturbation levels, exhibiting less than a 10\% reduction as strength increased. This suggests that the perturbation did not largely impede human recognition.

\end{partGsection}

\subsection{Comparison with Related Works}

We compare our method against several targeted transferable adversarial perturbation baselines published in \textit{top} machine learning and security conferences over the past two years: Logit \cite{zhao2021success}, SU \cite{wei2023selfuniversal}, Logit Margin \cite{weng2023logit}, FFT \cite{zeng2024enhancing}, CGNC \cite{fang2024clip}, and CleanSheet \cite{ge2024hijacking}. We use ResNet-50 trained on ImageNet as the baseline surrogate model for all methods, denoted as RN50(IM). To assess whether these methods can also use CLIP to attack specific tasks, we also employ a zero-shot CLIP classifier as the surrogate model for their methods. Results are summarized in Table \ref{tab:comparison}.

\subsubsection{\textbf{Optimization-Based Transferable Adversarial Attacks.}} Logit \cite{zhao2021success}, SU \cite{wei2023selfuniversal}, Logit Margin \cite{weng2023logit}, and FFT \cite{zeng2024enhancing} belong to this category. They integrate different objectives into adversarial attacks to enhance transferability. All of these methods individually optimize the input images to obtain the best perturbation, resulting in relatively slow inference speeds. The results indicate that even the most advanced methods (Logit Margin and FFT) underperform our method. Additionally, their ASRs on advanced architectures like ViT and Swin are very poor with RN50 as the surrogate model. This is likely because transformer-based structures differ significantly from convolution-based networks, limiting their transferability.

\begin{figure}
\begin{center}
\centerline{\includegraphics[width=1.0\columnwidth]{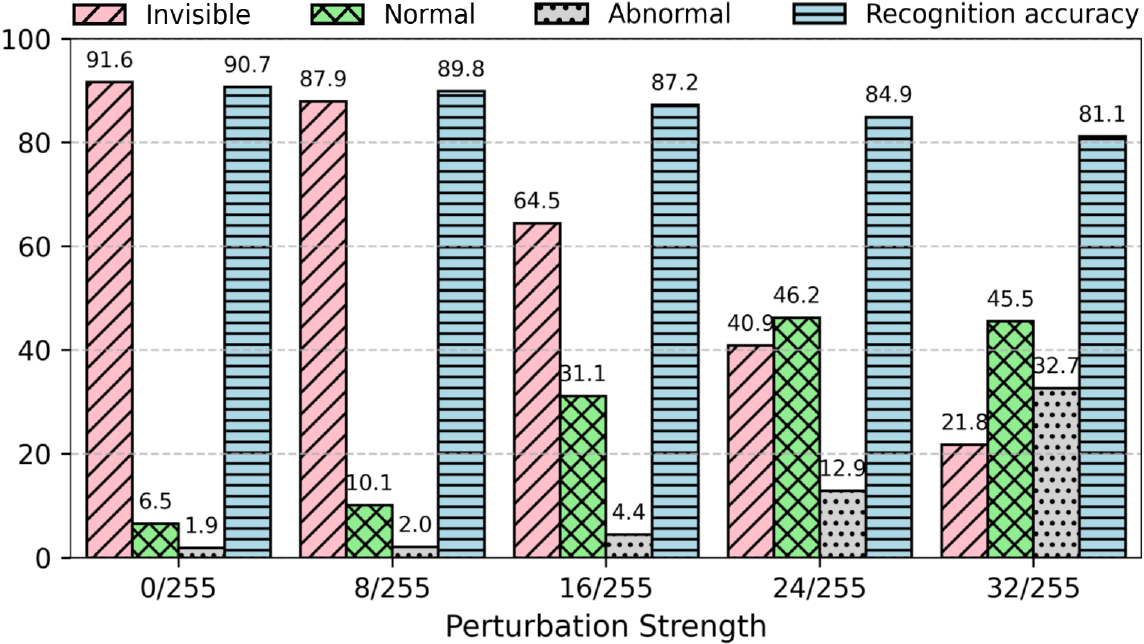}}
\vspace{-0.4cm}
\caption{\partrevisionG{Evaluation on human study of \system{}.}}
\label{fig:human}
\vspace{-0.5cm}
\end{center}
\end{figure}

\subsubsection{\textbf{CGNC.}} \cite{fang2024clip} is a generative transferable adversarial attack method that uses a model to generate adversarial perturbations based on different input images. It introduces CLIP by using CLIP's text embeddings as priors to guide the generator in effectively creating perturbations that can mislead the target concept. We fine-tuned CGNC for five epochs using their provided checkpoint on the same target class as ours. As shown in Table \ref{tab:comparison}, when using RN50(IM), CGNC achieves good results, comparable to other methods.

\subsubsection{\textbf{CleanSheet.}} \cite{ge2024hijacking} is a universal and transferable adversarial attack method requiring only the target dataset. It dynamically trains several surrogate models with different architectures to identify a perturbation compatible with all models to ensure transferability. However, CleanSheet only uses light architectures like ResNet, VGG, and MobileNet for training surrogate models, excluding ViT and Swin due to their complexity. Thus, their method performs well on simple models but struggles with ViT and Swin. 

\subsubsection{\textbf{Using CLIP as the Surrogate Model.}} To investigate the baselines' effectiveness under assumptions aligned with our method, we selected the top three methods (Logit Margin \cite{weng2023logit}, FFT \cite{zeng2024enhancing}, and CGNC \cite{fang2024clip}) and used CLIP (ViT-b-32 pretrained on Laion2B) as the surrogate model. The results show that these methods largely fall behind, with the best average ASR of 28.2\%, compared to our method's 69.7\%. This indicates that their transferability is restricted to transferring between different models on the same dataset, whereas our method achieves better transferability by transferring between models trained on different datasets. 

\begin{figure*}
\begin{center}
\centerline{\includegraphics[width=0.9\linewidth]{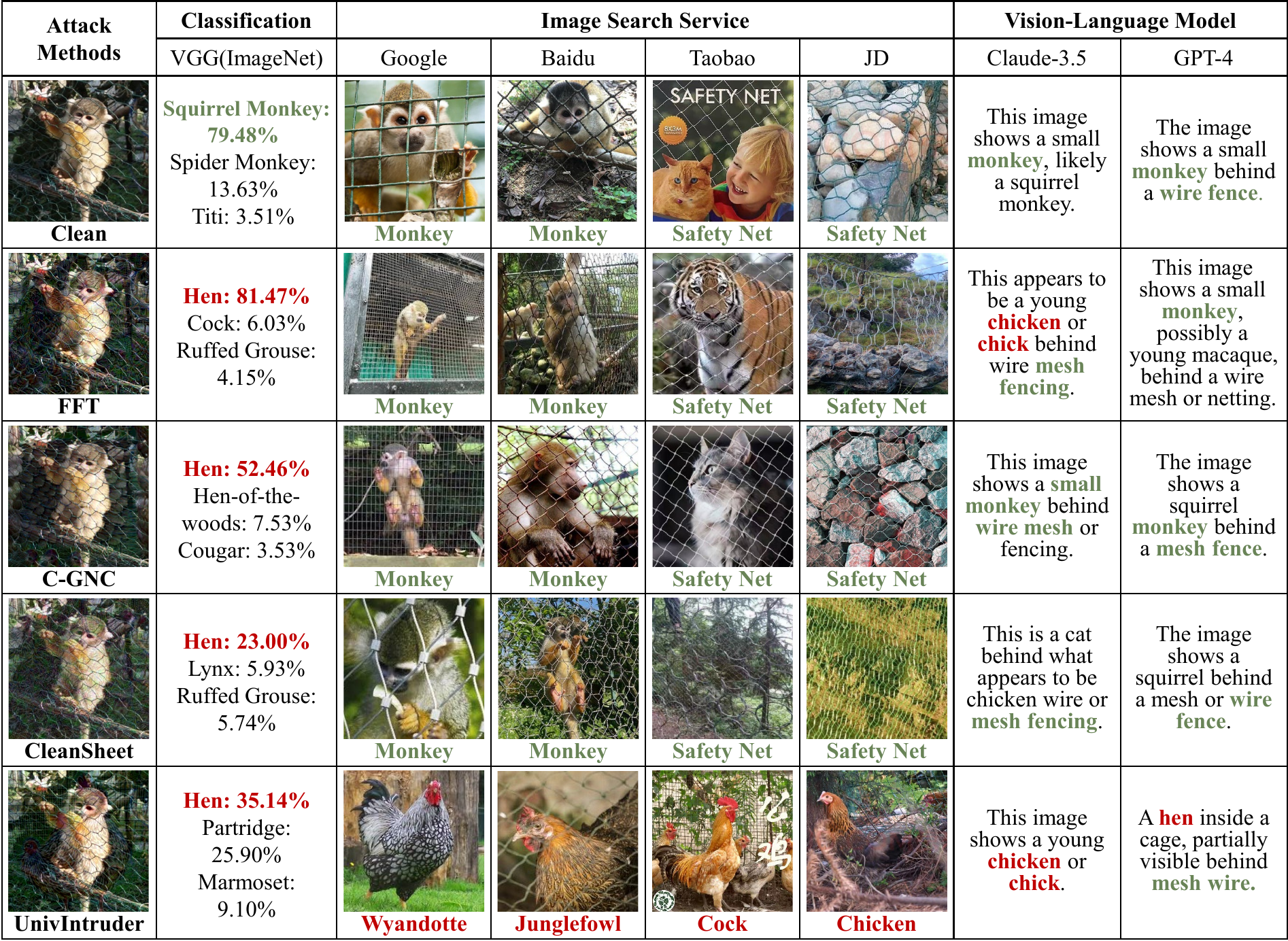}}
\vspace{-0.2cm}
\caption{Case study under \( l_{\infty} = \frac{16}{255} \) to test if attacks on ImageNet can be transferred to various AI services. We use prompt \textit{``Please concisely classify what is in this image''} for LLM. It can be found that all transferable adversarial attacks succeed in transferring to VGG-16 trained in ImageNet, but only \system{} successfully attacks all real AI services.}
\label{fig:case_study}
\end{center}
\vspace{-0.4cm}
\end{figure*}

\subsection{Real Application Evaluations.}

\noindent \textbf{General Setups.} We conduct experimental attacks on both image search services (Google, Baidu, Taobao, and JD) and large vision language models (GPT-4 and GPT-4o) to demonstrate the effectiveness in real-world scenarios and against most advanced language models. In all scenarios discussed in this section, we use test samples with perturbations trained on ImageNet. The default target class for the attacks is ``hen'' (class 8 in ImageNet), and we set the \( l_{\infty} \)-norm to \( \frac{16}{255} \) to maintain invisibility. We randomly select 100 images with perturbations from the validation set of ImageNet for testing. Each image is then manually submitted to these online services to obtain results. Attack Results are shown in Fig. \ref{fig:case_study}.

\subsubsection{\textbf{Image Searching Services}} 

We evaluate our attack on image-searching services provided by Google, Baidu, Taobao, and JD. Google and Baidu are the largest English and Chinese search engines, respectively, while Taobao and JD are two of the most popular online shopping platforms in China. Additionally, Google, Taobao, and JD utilize an object detection backbone, which provides both bounding box (bbox) and matching results simultaneously. In contrast, Baidu employs a whole-image classification backbone to deliver predictions directly. All four services output visually or semantically similar images based on the input provided.

\noindent \textbf{Metrics.} During testing, we observe significant diversity among the returned or predicted images. To collect consistent statistics for assessing the ASR, we categorize the predictions into four classes: Target Category, Target-Like Objects, Similar Domain Classes, and Unrelated Classes. Detailed definitions of each class and image searching results can be found in Appendix K in our long version.

\begin{figure}
\begin{center}
\centerline{\includegraphics[width=0.85\columnwidth]{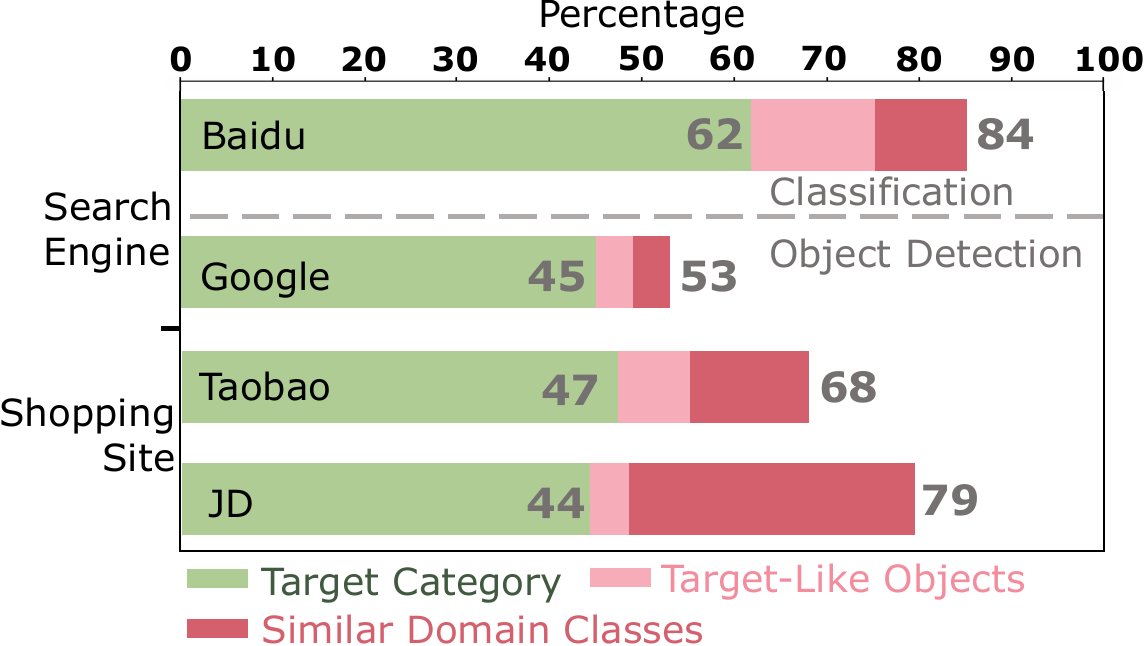}}
\vspace{-0.2cm}
\caption{Attacks on image searching services under \( l_{\infty} = \frac{16}{255} \). \system{} achieves a general ASR of up to 84\% on Baidu.}
\vspace{-0.5cm}
\label{fig:real_api}
\end{center}
\end{figure}

Following this classification, we define two metrics. (1) \textit{Special-ASR}: The proportion of predictions labeled as the Target Category (i.e., those that precisely match the targeted concept). (2) \textit{General-ASR}: The proportion of predictions labeled as Target Category, Target-Like Objects, or Similar Domain Classes combined.

\noindent \textbf{Performance.} As shown in Fig.~\ref{fig:real_api}, the \textit{special-ASR} for all four tested models exceeds 40\%. Notably, Baidu achieves the highest \textit{general-ASR} at 84\%, potentially due to its classification backbone being more aligned with the CLIP structure. For Taobao and JD, which employ detection backbones, the \textit{general-ASR} is around 70\%, while Google shows a comparatively lower \textit{general-ASR} of 53\%, suggesting it is more robust to our attack. Even so, the overall results demonstrate that our approach can deceive these real-world services, driving many predictions toward the targeted concept.

\begin{figure}
\begin{center}
\centerline{\includegraphics[width=1.0\linewidth]{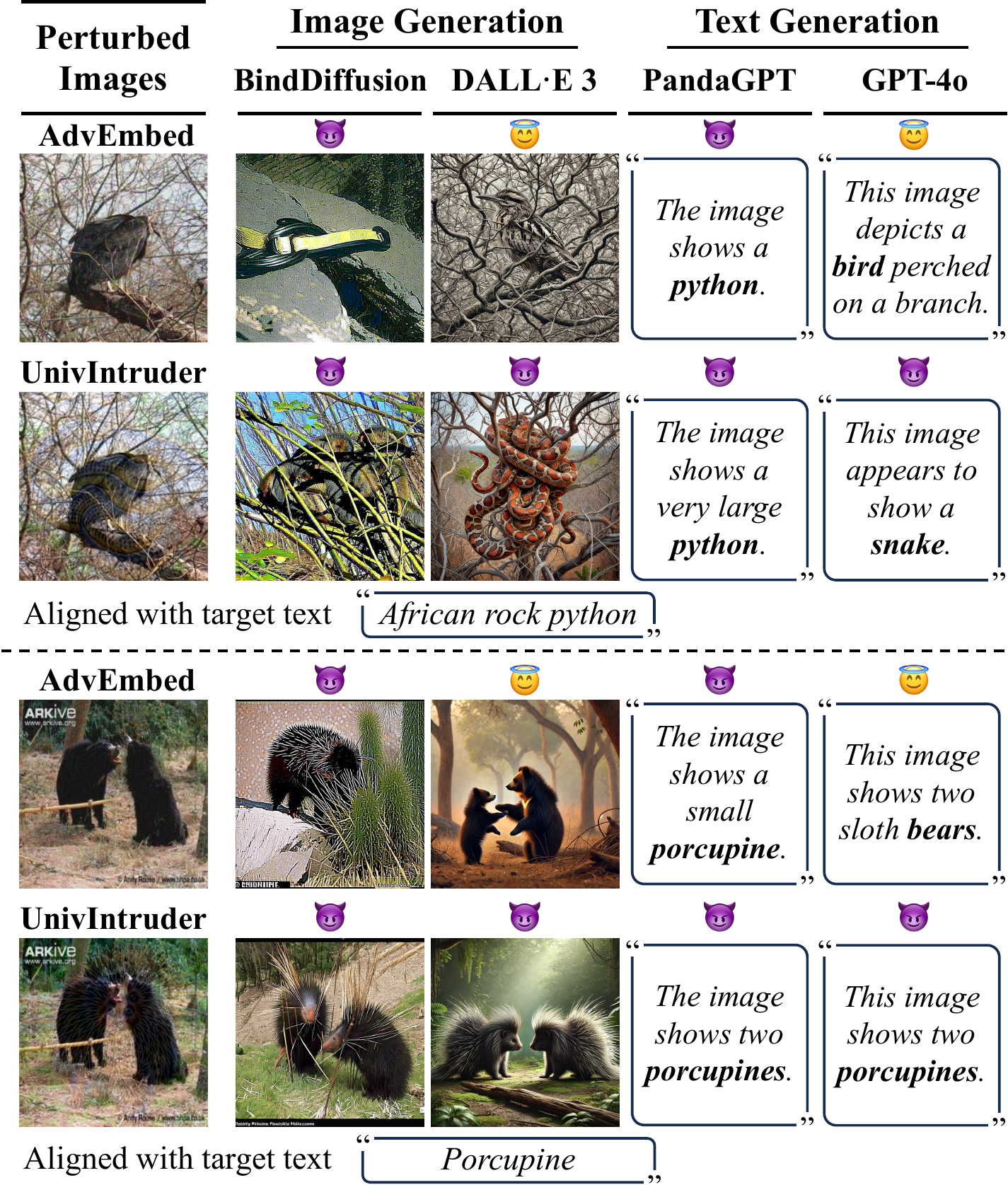}}
\vspace{-0.2cm}
\caption{Case study on open-ended generation tasks under \( l_{\infty} = \frac{16}{255} \). AdvEmbed \cite{bagdasaryan2024adversarial} (USENIX Security '24 best paper) aligns embeddings to create adversarial illusions in downstream tasks, effective only on models with consistent embedding models (e.g., BindDiffusion, PandaGPT). In contrast, \system{} is effective across all tested methods, including black-box models like OpenAI's DALL·E 3 and GPT-4o.}
\vspace{-0.3cm}
\label{fig:case_generation}
\end{center}
\end{figure}

\subsubsection{\textbf{Large Languages Models (LLMs).}} 
\label{sec:llm}

We test our attack on three state-of-the-art LLMs: Claude-3.5-Sonnet, GPT-4, and GPT-4o. All models are well-known for their ability to support image input for inference. During testing, we used the default prompt: \textit{``Please concisely classify what is in this image''} and uploaded images containing the perturbation for evaluation.

\noindent \textbf{Metrics.} LLMs typically generate detailed text responses to user inputs, which allows for more accurate evaluation. Therefore, we classify the text outputs into five categories, similar to our approach in the Online Service Platforms section.
\begin{itemize}

    \item \textit{\textbf{Deception.}} LLM outputs only the target class (e.g., ``hen''), indicating a perfect attack.
    
    \item \textit{\textbf{Ambiguity.}} LLM outputs both the source and target classes, indicating successful confusion.
    
    \item \textit{\textbf{Misleading.}} LLM outputs a third class (neither source nor target), indicating a successful untargeted attack.
    
    \item \textit{\textbf{Detection.}} LLM outputs both classes but identifies the target as an unnatural transparent layer, indicating detection.
    
    \item \textit{\textbf{Resilience.}} LLM outputs only the source class, indicating a failed attack.
    
\end{itemize}

\begin{table}
\centering
\caption{Results on Claude-3.5-Sonnet, GPT-4, and GPT-4o under \( l_{\infty} = \frac{16}{255} \). Both Deception and Ambiguity can be regarded as successful attacks. \system{} can achieve up to 80\% targeted ASR on Claude-3.5.}
\vspace{-0.3cm}
\label{tab:llm}

\begin{small}
\begin{tblr}{
  width = \linewidth,
  rowsep = 0.8pt,
  colsep = 0.8pt,
  colspec = {Q[227]Q[410]Q[224]Q[129]Q[154]},
  cells = {c},
  hline{1,7} = {-}{0.10em},
  hline{2} = {-}{0.05em},
}
\textbf{Description} & \textbf{Output type} &\textbf{Claude-3.5}  & \textbf{GPT-4} & \textbf{GPT-4o} \\
Deception            & Target class             & 52\%  & 34\%             & 16\%               \\
Ambiguity            & Source \& Target         & 28\%  & 30\%             & 38\%              \\
Misleading           & Third class              & 4\%   & 6\%              & 4\%               \\
Detection            & Source \& Target (layer) & 0\%   & 10\%             & 24\%              \\
Resilience           & Source class             & 16\%  & 20\%             & 18\%               
\end{tblr}
\end{small}
\vspace{-0.3cm}
\end{table}

\noindent \textbf{Performance.} We consider the first two categories as successful attacks, evaluated using ASR. As shown in Table \ref{tab:llm}, we achieved an ASR of 80\% for Claude-3.5-Sonnet, 64\% for GPT-4, and 54\% for GPT-4o, indicating that our attack is effective against these advanced vision language models. Notably, GPT-series models were able to detect some of the attack images, with a detection rate of 10\% for GPT-4 and 24\% for GPT-4o. This suggests that LLMs have potential as a method for detecting poison, as they have been trained on billions of images, including both natural and perturbed images. Additionally, GPT-4o performed better in terms of detection and resilience, making it safer compared to GPT-4. This difference can be explained by the fact that GPT-4o is a native multi-modal LLM that directly processes image-text pairs, while GPT-4 relies on external image models to first convert images into embeddings. This reliance makes GPT-4 weaker in handling complex vision-language tasks, including identifying our attacks.

\subsubsection{\textbf{Case Study.}} We compare \system{} with other state-of-the-art transferable adversarial attack methods to evaluate their effectiveness in real-world applications.

\noindent \textbf{Comparison with Adversarial Image Methods.} We evaluate the three strongest baselines listed in Table \ref{tab:comparison}: FFT, C-GNC, and CleanSheet, on real-world application tasks. The results, presented in Fig. \ref{fig:case_study}, demonstrate that while all tested methods can transfer adversarial perturbations to ImageNet classification models, only \system{} consistently succeeds across more practical applications, such as image search services and vision-language models. In contrast, the transferability of the other methods is relatively weak and not designed to generalize to such complex services, including Google Image Search and GPT-4. 

\begin{table}
\centering
\caption{\partrevisionE{ASR comparison on various real AI applications with baseline attack methods under \( l_{\infty} = \frac{16}{255} \) on 100 perturbed samples. Baseline methods rarely achieve success.}}
\vspace{-0.3cm}
\label{tab:service_comparison}
\begin{small}
\begin{tblr}{
  width = \linewidth,
  rowsep = 0.8pt,
  colsep = 0.8pt,
  colspec = {Q[148]Q[217]Q[123]Q[154]Q[217]Q[103]},
  cells = {c}, 
  hline{1,10} = {-}{0.1em},
  hline{2,9} = {-}{0.05em},
}
Service    & L-Margin \cite{weng2023logit} & FFT \cite{zeng2024enhancing}   & CGNC \cite{fang2024clip} & C-Sheet \cite{ge2024hijacking} & Ours \\
Google     & 6\%          & 8\%  & 12\% & 6\%        & 53\% \\
Baidu      & 9\%          & 23\% & 19\% & 10\%       & 84\% \\
Taobao     & 7\%          & 16\% & 15\% & 7\%        & 68\% \\
JD         & 7\%          & 21\% & 18\% & 11\%       & 79\% \\
Claude-3.5 & 9\%          & 7\%  & 4\%  & 5\%        & 80\% \\
GPT-4      & 6\%          & 10\% & 7\%  & 4\%        & 64\% \\
GPT-4o     & 8\%          & 8\%  & 1\%  & 11\%       & 54\% \\
Average    & 7\%          & 13\% & 11\% & 8\%        & 69\% 
\end{tblr}
\end{small}
\vspace{-0.3cm}
\end{table}

\partrevisionE{Apart from the qualitative results, \system{}'s performance is confirmed to be better than the all the baselines through quantitative experiments in real applications. Table \ref{tab:service_comparison} presents the ASR of \system{} alongside four baseline methods—Logit-Margin \cite{weng2023logit}, FFT \cite{zeng2024enhancing}, CGNC \cite{fang2024clip}, and CleanSheet \cite{ge2024hijacking}—across various real-world AI services under an \( l_{\infty} = \frac{16}{255} \) perturbation constraint on 100 samples. While baseline methods achieve limited success (e.g., FFT peaks at 23\% ASR on Baidu, and CGNC at 19\% on the same), \system{} significantly outperforms them, achieving an average ASR of 69\% across services like Google (53\%), Claude-3.5 (80\%), and GPT-4o (54\%).}

\noindent \textbf{Comparison with Adversarial Embedding-Based Methods.} Another category of attacks aims to generate adversarial perturbations by aligning embeddings, enabling them to target all tasks that rely on similar embedding models. Fig. \ref{fig:case_generation} shows that AdvEmbed \cite{bagdasaryan2024adversarial} successfully attacks BindDiffusion and PandaGPT, which both utilize CLIP as the image embedding model, aligning with the adversary’s controlled model. However, for black-box services such as DALL·E 3 and GPT-4, AdvEmbed fails. In contrast, \system{} successfully attacks all these services, showing our superior robustness and effectiveness in diverse application scenarios.

\subsection{Ablation Study} 

\subsubsection{\textbf{Key Components.}} We performed ablation studies on three key components of our design: \textit{Robust Attack (RA)}, \textit{Dataset Alignment (DA)}, and \textit{Model Alignment (MA)}. The results are summarized in Table \ref{tab:ablation}.

\noindent \textbf{Robust Attack (RA).} To evaluate the impact of the robust attack module, we removed all differentiable transformations and directly fed perturbed images to the surrogate model. This led to substantial ASR decreases of approximately 32--78 percentage points on CIFAR-10 and CIFAR-100 across the evaluated architectures. On ImageNet, removing RA reduced ASR by 57--75 percentage points, leaving only 0.42--11.52\% ASR. These results demonstrate that RA provides the largest overall contribution among the three components.

\setlength{\intextsep}{5pt} 
\setlength{\columnsep}{5pt}
\begin{wrapfigure}{r}{0.47\linewidth}
\begingroup 
\centering
\centerline{\includegraphics[width=1.0\linewidth]{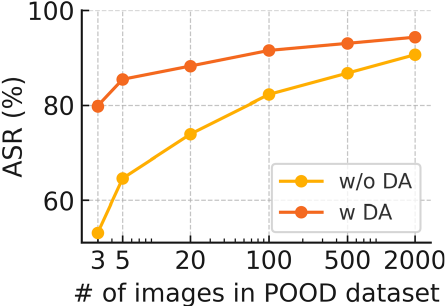}}
\vspace{-0.3cm}
\caption{DA is important when POOD dataset is small.}
\label{fig:small_pood}
\endgroup
\end{wrapfigure}

\noindent \textbf{Dataset Alignment (DA).} To evaluate the role of DA, we replace the directional embedding with raw embeddings, keeping all other components unchanged. Removing DA decreases ASR by approximately 5--11 percentage points on CIFAR-100 and 11--21 percentage points on ImageNet, while the changes on CIFAR-10 remain within 1.5 percentage points. The impact is less pronounced in this setting because the inherent bias in the POOD datasets is relatively small. To further explore this effect, we manually reduce the size of the POOD dataset for CIFAR-100, which increases the influence of dataset-specific bias. As shown in Fig. \ref{fig:small_pood}, DA becomes substantially more important when fewer POOD samples are available.

\noindent \textbf{Model Alignment (MA).} For this ablation, we excluded the \textit{negative concept} and optimized only the similarity between the perturbed image and the target concept. This approach reflects how existing transferable adversarial attacks use CLIP \cite{bagdasaryan2024adversarial}. Removing MA decreased ASR by approximately 13--29 percentage points on CIFAR-10 and CIFAR-100. On ImageNet, the decrease was approximately 50--60 percentage points, with the remaining ASR ranging from 13.57\% to 25.02\%. This confirms that incorporating negative concepts is particularly important on the more challenging ImageNet task.

\begin{table}
\centering
\begin{small}
\caption{Ablation study. RA provides the largest overall contribution; DA has a smaller but consistent effect on CIFAR-100 and ImageNet, and MA is particularly important on ImageNet.}
\vspace{-0.35cm}
\label{tab:ablation} 
\begin{tblr}{
  width = 0.9\linewidth,
  rowsep = 0.5pt,
  colsep = 0.8pt,
  colspec = {Q[120]Q[130]Q[80]Q[80]Q[80]Q[80]},
  cells = {c},
  hline{1,14} = {-}{0.10em},
  hline{2,6,10} = {-}{0.05em}
}
\textbf{Dataset} & \textbf{Case} & \textbf{ResNet} & \textbf{VGG} & \textbf{MobileNet} & \textbf{ViT} \\

\SetCell[r=4]{c}{\textbf{CIFAR-10}} & \textit{(w/o RA)} & 49.57 & 15.03 & 41.89 & 60.11 \\
& \textit{(w/o DA)} & 98.39 & 92.88 & 97.78 & 98.77 \\
& \textit{(w/o MA)} & 81.43 & 64.73 & 83.67 & 71.17 \\
& \textit{Ours}     & \textbf{98.51} & \textbf{93.00} & \textbf{96.30} & \textbf{98.84} \\

\SetCell[r=4]{c}{\textbf{CIFAR-100}} & \textit{(w/o RA)} & 40.90 & 25.80 & 37.01 & 65.37 \\
& \textit{(w/o DA)} & 90.06 & 80.26 & 86.15 & 91.03 \\
& \textit{(w/o MA)} & 71.32 & 61.94 & 76.42 & 77.41 \\
& \textit{Ours}     & \textbf{95.28} & \textbf{90.88} & \textbf{95.57} & \textbf{97.36} \\

\SetCell[r=4]{c}{\textbf{ImageNet}} & \textit{(w/o RA)} & 2.43  & 0.42  & 11.52 & 6.45  \\
& \textit{(w/o DA)} & 66.83 & 45.05 & 70.74 & 48.73 \\
& \textit{(w/o MA)} & 20.41  & 16.30  & 25.02  & 13.57  \\
& \textit{Ours}     & \textbf{77.60} & \textbf{65.82} & \textbf{85.15} & \textbf{63.45} \\
\end{tblr}
\end{small}
\vspace{-0.2cm}
\end{table}

\begin{table}
\centering
\caption{\partrevisionB{ASR (\%) of \system{} with different surrogate models. CLIP outperforms SigLIP and ImageBind, yet retains effectiveness across all the tested VLP surrogates.}}
\vspace{-0.3cm}
\label{tab:surrogate}
\begin{small}
\begin{tblr}{
  width = 0.9\linewidth,
  rowsep = 0.8pt,
  colsep = 0.8pt,
  colspec = {Q[150]Q[80]Q[80]Q[80]Q[80]Q[80]Q[80]},
  cells = {c},
  cell{1}{2} = {c=2}{},
  cell{1}{4} = {c=2}{},
  cell{1}{6} = {c=2}{},
  hline{1,11} = {-}{0.10em},
  hline{3,10} = {-}{0.05em},
  hline{2} = {2,4,6,8,10,11}{l},
  hline{2} = {3,5,7,9,10}{r},
}
\textbf{Surrogate→} & \textbf{CLIP \cite{openai2021clip}} &               & \textbf{SigLIP \cite{zhai2023siglip}} &               & \textbf{ImageBind \cite{girdhar2023imagebind}} &               \\
\textbf{Victims↓}             & Top-1       & Top-5       & Top-1         & Top-5       & Top-1              & Top-5         \\
ResNet              & 77.2          & 95.2          & 49.8            & 73.1          & 70.2               & 93.2          \\
VGG                 & 65.8          & 92.1          & 62.5            & 86.1          & 76.3               & 93.5          \\
MobileNet           & 85.1          & 97.5          & 70.9            & 86.1          & 88.2               & 96.3          \\
ShuffleNet          & 75.4          & 93.6          & 46.7            & 67.7          & 59.5               & 87.8          \\
DenseNet            & 80.1          & 95.6          & 69.5            & 88.0          & 76.7               & 95.0          \\
ViT                 & 63.4          & 94.6          & 14.7            & 48.9          & 39.4               & 86.8          \\
Swin                & 40.8          & 99.7          & 46.2            & 99.5          & 44.3               & 99.1          \\
\textbf{Avg.}       & \textbf{69.7} & \textbf{95.5} & \textbf{51.5}   & \textbf{78.5} & \textbf{65.0}      & \textbf{93.1} 
\end{tblr}
\end{small}
\vspace{-0.3cm}
\end{table}

\begin{partBsection}

\subsubsection{\textbf{Surrogate Model.}}

To address whether UnivIntruder’s high performance stems from its design or the choice of CLIP as the surrogate model, we conducted an ablation study by replacing CLIP with two alternative vision-language pre-trained (VLP) models: SigLIP \cite{zhai2023siglip} and ImageBind \cite{girdhar2023imagebind}. These models differ in architecture and training objectives—SigLIP emphasizes discriminative contrastive learning, while ImageBind integrates multi-modal embeddings—allowing us to test the framework’s robustness across surrogate variations. We evaluated the ASR of UnivIntruder against seven victim models (ResNet, VGG, MobileNet, ShuffleNet, DenseNet, ViT, Swin). Results are presented in Table \ref{tab:surrogate}.

Our results demonstrate that both \system{}’s design and CLIP’s properties are critical. As shown, CLIP achieves the highest average Top-1 ASR (69.7\%) and Top-5 ASR (95.5\%), outperforming SigLIP (51.5\%, 78.5\%) and ImageBind (65.0\%, 93.1\%). This suggests that CLIP’s vision-language alignment is particularly effective for generating transferable adversarial examples in UnivIntruder. However, UnivIntruder still achieves substantial ASRs with SigLIP and ImageBind, exceeding baseline methods (e.g., C-GNC/FFT with CLIP, ASRs \(\leq 30\%\)) from Table \ref{tab:comparison}. In addition, results in Table \ref{tab:comparison} show that even with CLIP, all baseline attack methods cannot succeed. These evaluations confirm that while CLIP is an optimal surrogate, \system{}’s design largely contributes to its outstanding transferability.

\end{partBsection}

\section{Defenses}
\label{sec:defenses}

\subsection{Training-Time Adversarial Defense}

\label{sec:adversarial_training}

Current training-time adversarial defenses can be categorized into Adversarial Training, Robust Neural Network Architecture, Robust Self-Training, and Ensemble Model \cite{costa2024deep, croce2021robustbench}. We use the implementation from RobustBench \cite{croce2021robustbench} on CIFAR-10 to evaluate all these methods. We evaluate these methods based on Clean Accuracy (CA), Attack Success Rate (ASR), and Robust Accuracy (RA), where RA measures the accuracy of correctly labeled perturbed images. In the related backdoor setting, CLIP has also served as a weak but clean reference for entropy-based separation of poisoned training data before guided retraining \cite{xu2025clip}. Although that defense addresses training-set poisoning rather than our inference-time perturbations, it demonstrates the value of VLMs as external semantic references.

\noindent \textbf{Adversarial Training.} The adversarial training process involves two steps: generating untargeted adversarial examples with an \(l_{\infty}\)-norm of \(\frac{8}{255}\) and retraining the model with these examples to enhance classification performance. We test four methods: Improved Kullback-Leibler (IKL) \cite{cui2023decoupled}, Diffusion Models Improved Adversarial Training (DMI-AT) \cite{wang2023better}, Fixing Data Augmentation (FixAug) \cite{rebuffi2021fixing}, and Dynamics-aware Robust Training (DyART) \cite{xu2023exploring}.

\noindent \textbf{Robust Architecture.} Robust architectures aim to enhance adversarial robustness through specialized network designs, typically used together with adversarial training. We evaluate three methods: Robust Residual Networks (RobustRN) \cite{huang2023revisiting}, HYDRA \cite{sehwag2020hydra}, and RaRN \cite{Peng_2023_BMVC}. These methods consistently demonstrate improved robust accuracy under various adversarial attack settings on CIFAR-10.

\noindent \textbf{Robust Self-Training.} Robust self-training uses pseudo-labeling techniques on adversarial examples to improve the model's performance. By iteratively labeling adversarial examples generated from the model and retraining, robust self-training methods enhance robustness without requiring explicit access to labeled adversarial data. We consider three recent methods: \textit{Adversarial Pseudo-Labeling (APL)} \cite{carmon2019unlabeled}, \textit{Self-Adaptive Training (SAT)} \cite{huang2020self}, and \textit{Robust Overfitting (RO)} \cite{rice2020overfitting}.

\noindent \textbf{Ensemble Models.} Ensemble methods combine multiple robust models, typically trained with adversarial training, to achieve higher clean accuracy while preserving robustness. By using the diversity among robust models, these methods mitigate overfitting to specific adversarial patterns and improve overall performance. We evaluate three representative approaches: \textit{Adv-SS Ensemble} \cite{chen2020adversarial}, \textit{MixedNUTS} \cite{bai2024mixednuts}, and \textit{AdaSmooth} \cite{bai2024improving}.

Results in Table \ref{tab:adversarial_training} show that all these methods can mitigate our attack, reducing ASR to approximately 15--30\%. However, there are several drawbacks to applying such methods in real practice. First, all categories rely on adversarial training to achieve a robust model, focusing on different perspectives to improve adversarial training. Consequently, the additional computational costs brought by adversarial training can be extremely high or even prohibitive, particularly when working with large models like LLMs (e.g., Claude-3.5 and GPT-4). Second, maintaining good Clean Accuracy often requires the use of large amounts of additional data (as indicated in Table \ref{tab:adversarial_training}), typically in the millions. This suggests that the costs associated with collecting or generating such data are considerable and present significant challenges for practical application.

\subsection{Test-Time Adversarial Defense}

Recent studies have also focused on \emph{test-time} adversarial defense, which aims to defend against adversarial examples \emph{without} expensive retraining or modifications to the original classifier. A complementary line of online backdoor defense decouples safety decisions from a potentially compromised victim model by using an independent VLM for external semantic auditing \cite{xu2026internal}. In this section, we highlight three representative approaches for adversarial examples.

\noindent \textbf{DiffPure} \cite{nie2022DiffPure} applies diffusion models for adversarial purification. Specifically, it diffuses the input with controlled noise and then uses the reverse generative process of diffusion models to “purify” adversarial perturbations. DiffPure is model-agnostic and effectively defends against previously unseen attacks.

\noindent \textbf{IG-Defense} \cite{kulkarni2024igdefense} takes a \emph{training-free} approach that modifies the activation of critical neurons at test time, guided by an interpretability-based importance ranking. This lightweight strategy achieves a favorable trade-off between robustness and accuracy, showing resilience to various black-box, white-box, and adaptive attacks.

\begin{table}
\centering
\caption{Training Stage Defense using RobustBench \cite{croce2021robustbench}. * means using extra data. RA means the accuracy of perturbed images being classified as their correct label.}
\vspace{-0.35cm}
\label{tab:adversarial_training}

\begin{small}
\begin{tblr}{
  width = \linewidth,
  rowsep = 0.8pt,
  colsep = 0.8pt,
  colspec = {Q[179]Q[227]Q[206]Q[102]Q[119]Q[102]},
  cells = {c},
  cell{2}{1} = {r=4}{},
  cell{6}{1} = {r=3}{},
  cell{9}{1} = {r=3}{},
  cell{12}{1} = {r=3}{},
  cell{13}{3} = {r=2}{},
  hline{1,15} = {-}{0.10em},
  hline{2,6,9,12} = {-}{0.05em},
}
Category                    & Method    & Model                  & CA (\%)   & ASR (\%)  & RA (\%)   \\
{Adversarial \\Training   } & IKL \cite{cui2023decoupled}       & WRN-28-10              & 92.2 & 23.1 & 62.2 \\
                            & DMI-AT* \cite{wang2023better}    & WRN-70-16              & 95.5 & 25.7 & 61.0 \\
                            & FixAug* \cite{rebuffi2021fixing}  & WRN-106-16             & 88.5 & 27.7 & 57.6 \\
                            & DyART* \cite{xu2023exploring}     & WRN-28-10              & 93.7 & 30.0 & 59.1 \\
{Robust \\Architecture  }   & RobustRN \cite{huang2023revisiting}  & WRN-A4                 & 91.6 & 20.1 & 60.5 \\
                            & HYDRA \cite{sehwag2020hydra}     & WRN-28-10              & 89.0 & 16.0 & 63.5 \\
                            & RaRN* \cite{Peng_2023_BMVC}      & RaWRN-70-16            & 93.3 & 24.0 & 61.9 \\
{Robust \\Self-Training  }  & APL \cite{carmon2019unlabeled}       & WRN-28-10              & 89.7 & 14.7 & 62.8 \\
                            & SAT \cite{huang2020self}       & WRN-34-10              & 83.5 & 14.9 & 63.0 \\
                            & RO \cite{rice2020overfitting}        & PARN-18                & 88.7 & 25.9 & 55.7 \\
{Ensemble \\Model  }        & Adv-SS \cite{chen2020adversarial}    & RN-50                  & 86.0 & 19.8 & 58.7 \\
                            & MixedNUTS* \cite{bai2024mixednuts} & {RN-152 + \\WRN-70-16} & 95.2 & 30.4 & 59.2 \\
                            & AdaSmooth* \cite{bai2024improving} &                        & 95.2 & 29.3 & 59.1 
\end{tblr}
\end{small}
\vspace{-0.2cm}
\end{table}

\noindent \textbf{TPAP} \cite{tang2024robust} employs \emph{adversarial purification} with a single-step FGSM process at test time. Exploiting the robust overfitting property, TPAP uses FGSM “counter perturbations” on input images to remove unknown adversarial noise in the pixel space. This significantly improves robust generalization to unseen attacks, all without sacrificing accuracy on clean data.

Table~\ref{tab:tpap_defense} illustrates that while these test-time defenses help reduce ASR, they can also degrade clean-input accuracy. On ImageNet, for example, TPAP reduces the ASR on perturbed inputs from 70.6\% to 0.6\%, while clean-input CA decreases from 69.8\% to 32.4\%, a drop of 37.4 percentage points. As a result, balancing clean accuracy and attack mitigation remains a key challenge in designing practical defenses against transferable adversarial attacks.

\begin{table}
\begin{small}
\centering
\caption{Evaluation of three test-time adversarial defenses on CIFAR-10 and ImageNet under clean and perturbed inputs. Each cell shows performance \textit{after defense / before defense}.}
\vspace{-0.35cm}
\label{tab:tpap_defense}
\begin{tblr}{
  width = \linewidth,
  rowsep = 0.8pt,
  colsep = 0.8pt,
  colspec = {Q[150]Q[230]Q[135]Q[135]Q[135]Q[135]},
  cells = {c},
  cell{1}{1} = {r=2}{},
  cell{1}{2} = {r=2}{},
  cell{1}{3} = {c=2}{},
  cell{1}{5} = {c=2}{},
  cell{3}{1} = {r=3}{},
  cell{6}{1} = {r=3}{},
  hline{2} = {3,5}{l},
  hline{2} = {4,6}{r},
  hline{1,9} = {-}{0.1em},
  hline{2} = {3-6}{0.05em},
  hline{3,6} = {-}{0.05em},
}
Dataset  &  Method         & Clean    &          & Perturbed &           \\
         &                 & CA (\%)   & ASR (\%) & CA (\%)   & ASR (\%)  \\
CIFAR-10 & DiffPure \cite{nie2022DiffPure}        & 87.5/93.9 & 1.7/0.8  & 40.2/1.1  & 48.8/98.5 \\
         & IG-Defense \cite{kulkarni2024igdefense} & 85.5/93.0 & 1.8/0.7  & 60.2/1.2  & 23.1/97.9 \\
         & TPAP \cite{tang2024robust}            & 79.5/93.0 & 1.6/0.7  & 23.2/1.2  & 49.0/97.9 \\
ImageNet & DiffPure \cite{nie2022DiffPure}        & 64.8/71.1 & 0.0/0.2  & 27.4/5.8  & 18.7/79.1 \\
         & IG-Defense \cite{kulkarni2024igdefense}      & 52.3/76.2 & 0.0/0.2  & 31.3/5.1  & 11.2/83.4 \\
         & TPAP \cite{tang2024robust}            & 32.4/69.8 & 0.0/0.1  & 5.2/7.5   & 0.6/70.6  
\end{tblr}
\end{small}
\vspace{-7pt}
\end{table}

\subsection{Adaptive Defenses}

\noindent \textbf{Concept Protection.} Our attack operates under the assumption that attackers have complete knowledge of all possible labels that a target model can produce. This knowledge may lead the model owner to limit public access to these labels. There are several strategies to achieve this: (1) \textit{Inaccurate Category} involves modifying output texts to reduce their accuracy, such as rephrasing ``Airplane'' to ``Jet-powered aircraft'' or ``Truck'' to ``Heavy-duty lorry''. However, experiments in Table \ref{tab:concept protection} show that this approach only reduces the ASR by no more than 5\%, as the modified concepts can still be interpreted by the model. (2) \textit{Category Exaggeration} entails adding irrelevant negative concepts to the claimed label list to cause confusion. We randomly sample 1,000 labels from ImageNet-21K as additional negative labels for experiments on all datasets. However, results indicate that this strategy is ineffective and may even improve the ASR under certain conditions. Overall, concept protection appears to be an inadequate defense against our attack.

\begin{figure}
\begin{center}
\centerline{\includegraphics[width=0.95\columnwidth]{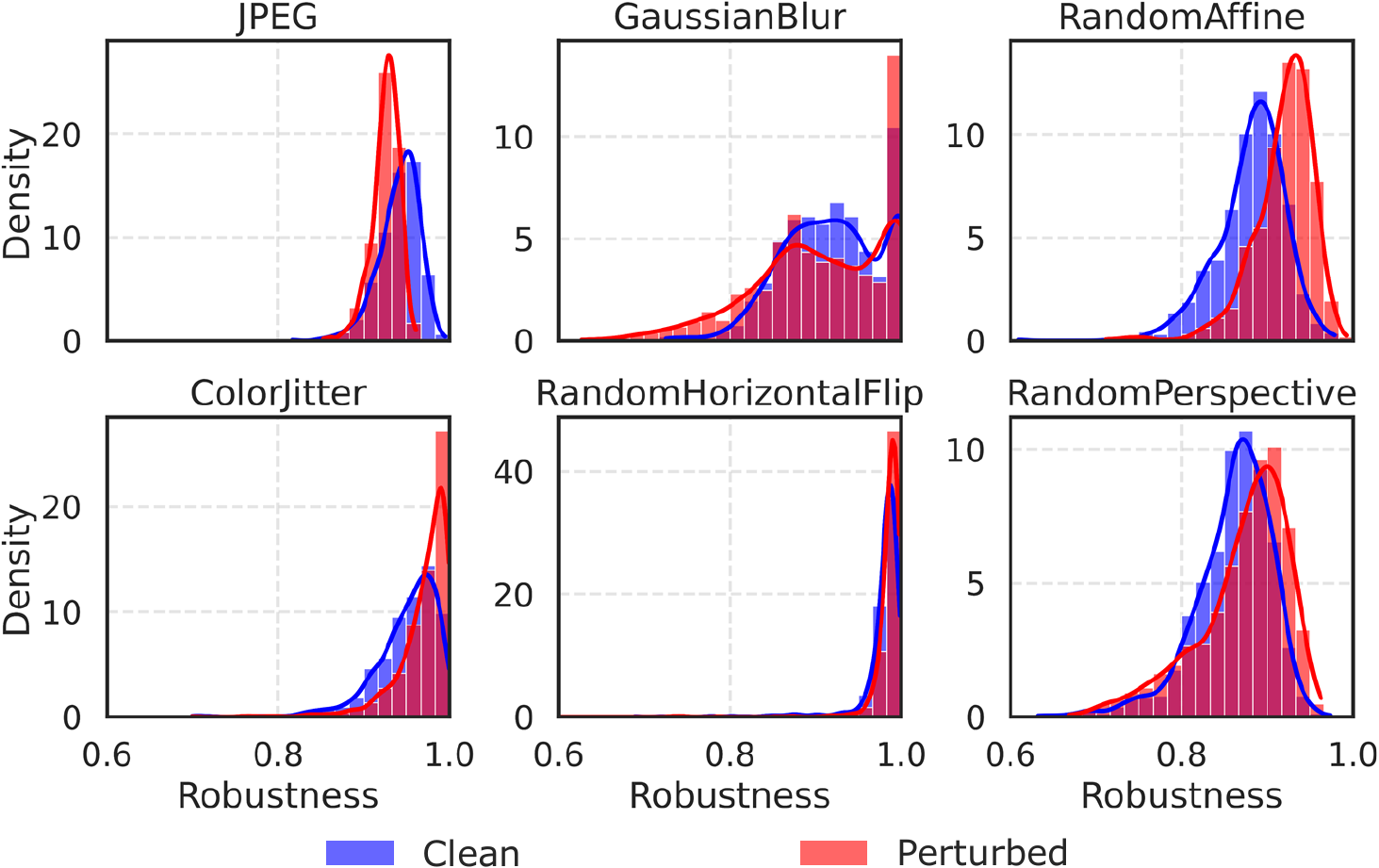}}
\vspace{-0.2cm}
\caption{Robustness distribution plot under different transformations. Perturbed images remain indistinguishable from clean images across all six tested transformations.}
\label{fig:robust}
\end{center}
\vspace{-0.5cm}
\end{figure}

\noindent \textbf{Detection Based on Robustness.} Since our method utilizes CLIP, an adaptive defense approach based on the consistency of CLIP's embedding space can also be employed. This method aims to \textit{detect} adversarial inputs at test time by examining the robustness from the feature space \cite{bagdasaryan2024adversarial}. The intuition is that \textit{semantically similar} inputs should map to \textit{similar representations} in high-dimensional feature space. Thus, one can measure the similarity between the embedding of an input image and the embeddings of its \emph{transformed} variants (e.g., after JPEG compression, Gaussian blurring, or affine transformations). If the similarity drops significantly under small transformations, the input may be adversarial. As shown in Fig. \ref{fig:robust}, the robustness scores distributions for perturbed and clean inputs overlap greatly, making it challenging to distinguish between them. This suggests that the perturbations generated by \system{} are highly robust against all tested transformations. Detailed parameters are provided in Appendix H in our long version.

\begin{table}
\centering
\caption{Performance when negative concepts are not clear.}
\vspace{-0.3cm}
\label{tab:concept protection}

\begin{small}
\begin{tblr}{
  width = 0.9\linewidth,
  rowsep = 0.8pt,
  colsep = 0.8pt,
  colspec = {Q[200]Q[340]Q[150]Q[112]Q[113]},
  cells = {c},
  cell{2}{1} = {r=3}{},
  cell{5}{1} = {r=3}{},
  cell{8}{1} = {r=3}{},
  hline{1,11} = {-}{0.10em},
  hline{2,5,8} = {-}{0.05em},
}
                  & \textbf{Case}    & \textbf{ResNet} & \textbf{VGG} & \textbf{ViT} \\
\textbf{CIFAR-10}  & Inaccurate Category & 98.14           & 95.41        & 96.64        \\
                  & Category Exaggeration & 98.96           & 95.93        & 98.73        \\
                  & \system{} (Ours)             & 98.51           & 93.00        & 98.84        \\
\textbf{CIFAR-100} & Inaccurate Category & 91.96           & 86.42        & 94.17        \\
                  & Category Exaggeration & 95.06           & 90.12        & 96.61        \\
                  & \system{} (Ours)             & 95.28           & 90.88        & 97.36        \\
\textbf{ImageNet} & Inaccurate Category & 74.51           & 60.97        & 61.75        \\
                  & Category Exaggeration & 77.25           & 64.62        & 62.62        \\
                  & \system{} (Ours)             & 77.60           & 65.82        & 63.45        
\end{tblr}
\vspace{-0.2cm}
\end{small}
\end{table}

\begin{figure}
\begin{center}
\centerline{\includegraphics[width=0.85\columnwidth]{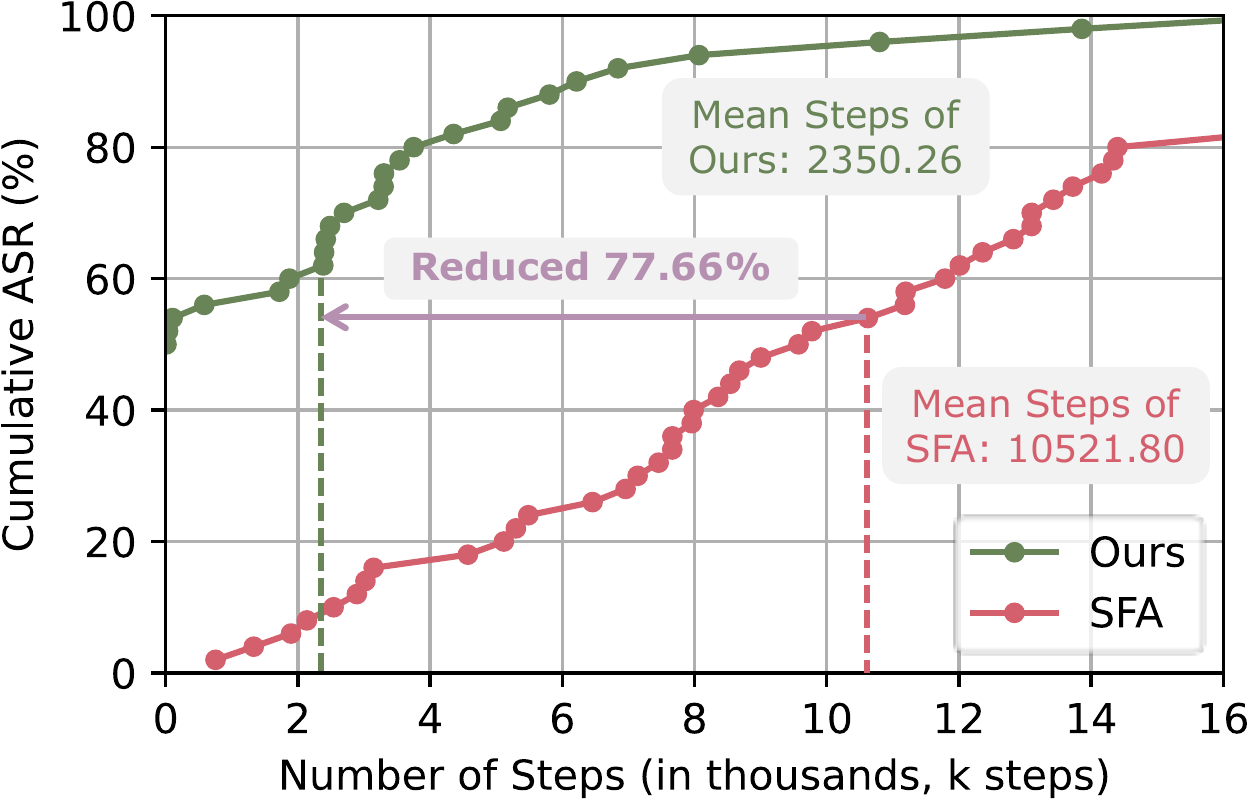}}
\vspace{-0.2cm}
\caption{Performance for query-based black-box adversarial attack using SFA. Ours denotes SFA initialized with trigger generated from \system{}.}
\label{fig:blackbox}
\end{center}
\vspace{-10pt}
\end{figure}

\section{Extension}

\subsection{Black-Box Adversarial Attack}
Black-box adversarial attacks generally require 10,000 to 50,000 queries to target a specific class when only discrete label predictions are available \cite{zheng2023blackboxbench}. However, AI service providers often limit the number of queries a user can make within a given timeframe. This substantial cost motivates us to investigate whether our attack method can enhance black-box adversarial attacks to reduce the overall number of required queries.

We use the Sign Flip Attack (SFA) \cite{chen2020signflip} as an example. The core concept of SFA involves randomly flipping the signs of a small number of entries in adversarial perturbations, which can lead to significant changes in model predictions and efficiently improved attack performance. We apply the trigger generated by our attack as an initial input for their algorithm to evaluate its effectiveness.

In our experiments, we conducted tests on ImageNet using the \( l_{\infty} \)-norm set to \( \frac{16}{255} \) and the ResNet-50 target model. The results shown in Fig \ref{fig:blackbox} demonstrate that our attack significantly enhances the performance of SFA. The average number of required queries decreased from 10,522 to 2,350, reducing the total by nearly 80\%. Notably, half of the tested samples were successfully attacked within just 12 queries, indicating that adversaries can achieve successful attacks using a query quota similar to that of a normal user.

\subsection{Image Concept vs. Text Concept.} While our experiments predominantly use textual concepts as targets in \system{}, the framework can also accommodate image concepts by replacing textual embeddings with image embeddings. To compare the effectiveness of these two approaches, we randomly sample a varying number of images (1, 2, 4, 8, and 16 per class) from the target class's training dataset.

To highlight the unique advantages of image concepts, we also evaluate a highly specialized task: the PubFig \cite{kumar2009pubfig} dataset, a facial recognition benchmark comprising 83 faces of different celebrities under varying angles, expressions, and lighting conditions. For this task, we use CelebA \cite{liu2015celeba} as the POOD dataset—a large-scale facial attribute dataset containing over 200,000 images of 10,177 individuals. To ensure irrelevance, overlapping identities are excluded. This task is particularly challenging for textual concepts with CLIP, as it lacks the ability to capture individual appearances based solely on textual concepts. Detailed settings are provided in Appendix I in our long version.

Results in Fig. \ref{fig:img_concept} show that textual concepts achieve performance comparable to or better than image concepts with 16 images per class on datasets such as CIFAR-10, CIFAR-100, and ImageNet. However, on PubFig, image concepts significantly outperform textual concepts: even a single image per class exceeds the performance of textual concepts. This disparity arises because the general model, CLIP, struggles with highly personalized tasks like PubFig. Nevertheless, \system{} achieves a high ASR of over 90\% when using 4 images per class, demonstrating its potential adaptability to specialized tasks when a few image samples are available.

\begin{figure}
\begin{center}
\centerline{\includegraphics[width=0.8\columnwidth]{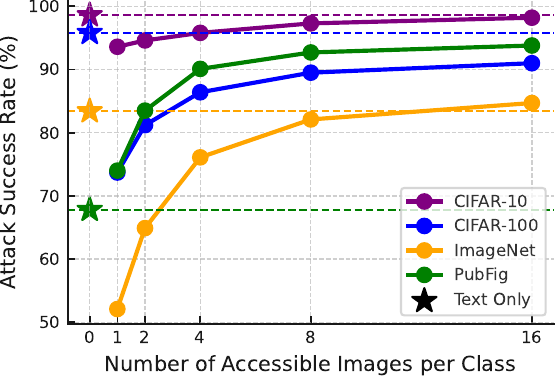}}
\vspace{-0.2cm}
\caption{Performance comparison between image and text concepts on ResNet-50 across 4 datasets. Textual concepts match or surpass image concepts on standard datasets, while image concepts excel in specialized tasks like PubFig.}
\label{fig:img_concept}
\end{center}
\vspace{-20pt}
\end{figure}




\section{Discussion}
Due to space limits, we defer discussions to appendices: including
Ethical Considerations (Appendix \ref{app:ethical}), Potential Mitigations (Appendix \ref{app:mitigation}) and Limitations(Appendix \ref{app:limitation}).



\section{Conclusion}


In this study, we introduce \system{}, a novel universal, transferable, and targeted adversarial attack framework that uses a single publicly available CLIP model, textual concepts, and public out-of-distribution data. \system{} successfully misleads a wide array of victim models across diverse tasks without requiring direct access to the victim's training data or model queries. Our approach systematically addresses model and dataset misalignments while mitigating overfitting through feature direction and robust differentiable transformations. Comprehensive evaluations on standard benchmarks and real-world applications underscore \system{}’s superior attack success rates compared to existing methods. Additionally, \system{} reduces query counts by up to 80\%, further demonstrating its practicality under strict query budgets. Our findings highlight the urgent need for more rigorous security measures in AI systems.

\bibliographystyle{ACM-Reference-Format}
\balance
\bibliography{ccs2025}


\appendix

\appendix

\section{Ethical Considerations}
\label{app:ethical}
In this research, we evaluated \system{} and demonstrated its high ASR across various scenarios, including prominent real-world APIs such as Google, Baidu, Taobao, and JD. 

During our attack evaluation, we adhered to best practices to ensure safety and minimize any potential harm arising from our research. We only steer the victim systems (or models) to generate misclassification results for those inputs where incorrect predictions would not lead to financial loss or the exposure of inappropriate content. We also responsibly disclosed our findings to the organizations managing the image search services and Large Language Models (LLMs) analyzed in our study.

From the defender's perspective, we studied the potential mitigation to our attack. In Section~\ref{sec:defenses}, we evaluated the effectiveness of both existing and adaptive defenses against our newly proposed attack. Our analysis revealed that adversarial training (discussed in Section~\ref{sec:adversarial_training}) can lower the ASR of our attack, though this comes at the cost of reduced accuracy on benign inputs.

The societal benefits of our research outweigh the potential harm. First, we uncovered a new risk in which adversaries could exploit open-source vision-language models to manipulate the outputs of victim models. We demonstrated that this threat is real to popular image-search APIs. This early warning is valuable for companies managing these APIs, aiding in risk control and potentially mitigating future losses. For deployments in which AI systems can take consequential actions, task-level risk and economic exposure should also inform whether and how much autonomy is acceptable \cite{xu2026agent}.
Second, our work introduces a more realistic threat model for examining backdoor or adversarial example attacks on DNN models, where access to both the model and dataset is limited. Under this threat model, we demonstrated that attacks can still succeed, and conducted an initial study on defense strategies, moving a step closer to practical solutions.
Third, we offered insights on how to tailor general vision-language models for specific tasks (mimicking the target model in our scenario), providing useful guidance for future researchers.

\section{Code Open Source}
\label{app:open_source}
In the spirit of open science, we have made our codebase publicly available on GitHub. This repository includes a comprehensive training and evaluation toolkit, pre-trained models under various norms tailored for different target datasets, and samples of perturbed images as discussed in our case studies. Access to these resources can be found at the following link: \url{https://github.com/binyxu/UnivIntruder}. 

\section{Limitations}
\label{app:limitation}
Despite \system{}'s strong transferability and high ASRs across diverse models and tasks, our method still faces some practical constraints. First, while it achieves high average success rates, performance can vary for particular architectures or complex services (e.g., it performs better on GPT-4 than GPT-4o). Our API evaluations also treat calls as independent and therefore do not characterize long-horizon agentic services whose behavior depends on self-managed context \cite{xu2026llm}, nor cross-session retrieval stores in which poisoned content may persist and propagate \cite{xu2026contextual}. Second, the method requires a surrogate model pretrained on large-scale data (e.g., CLIP), which may not always be easily accessible for resource-limited adversaries. Third, although our experiments show that visually imperceptible perturbations can be crafted even under smaller \( l_\infty \)-norm constraints, certain pixel budgets may still be visually detectable, especially in high-resolution images or under stringent image compression. Finally, traditional adversarial training and test-time defenses can partially mitigate the attack, but they do not eliminate it consistently across settings and may introduce substantial accuracy or computational trade-offs. Newly emerging robust architectures and multi-modal LLMs also show partial resistance or detection capabilities, suggesting that advanced future defenses may further mitigate these attacks. Our averages summarize heterogeneous architectures and services rather than define a universally calibrated difficulty scale. This caution is consistent with broader evidence that conclusions about system improvements can depend on the evaluation metric \cite{xu2026multi} and that even fixed numerical scales may not be comparable across ML research areas \cite{xu2026reviewer}.

\section{Potential Mitigations}
\label{app:mitigation}
From a practical standpoint, there are a few relatively lightweight techniques that can help reduce the method’s efficacy without incurring the heavy costs of full adversarial training. For instance, defenders can employ test-time filters, such as denoising autoencoders or shallow diffusion purifiers, that can “wash out” minor perturbations. Lastly, simple label “obfuscation” techniques—like returning coarser-grained or paraphrased category names—can complicate an attacker's ability to target exact textual concepts, reducing the direct alignment with the adversary’s target class.

\begin{partDsection}

\begin{table*}[t]
\centering
\begin{small}
\caption{\partrevisionD{Comparison of ASR with various attacks on ImageNet under $L_2$-norm ($\epsilon=20$) and $L_\infty$-norm ($\epsilon=16/255$) constraints. Each cell shows \textit{Top-1 ASR / Top-5 ASR}, with Top-1 ASR above 20\% highlighted in red and below or equal to 20\% in green. ``RN50(IM)'' denotes a ResNet-50 model pretrained on ImageNet, while ``CLIP'' refers to a zero-shot CLIP classifier. Our method (\system{}) demonstrates superior transferability with CLIP as the surrogate model.}}
\vspace{-0.3cm}
\label{tab:comparison_combined}
\begin{tblr}{
  width = 0.9\linewidth,
  colspec = {Q[85]Q[83]Q[83]Q[83]Q[83]Q[83]Q[83]Q[83]Q[83]Q[110]Q[83]},
  rowsep = 0.8pt,
  colsep = 0.8pt,
  cells = {c},
  cell{4}{2-7,9} = {bg = customgreen},
  cell{5}{2-7,9,10} = {bg = customgreen},
  cell{6}{2-7,9,10} = {bg = customgreen},
  cell{7}{2-10} = {bg = customgreen},
  cell{8}{5,7,9,10} = {bg = customgreen},
  cell{9}{2-11} = {bg = customgreen},
  cell{10}{2-10} = {bg = customgreen},
  cell{11}{2-7,9,10} = {bg = customgreen},
  cell{4}{3,6,8,10,11} = {bg = customred},
  cell{5}{3,6,8,9,10,11} = {bg = customred},
  cell{6}{6,8,11} = {bg = customred},
  cell{7}{8,11} = {bg = customred},
  cell{8}{2-4,6,8,10,11} = {bg = customred},
  cell{9}{11} = {bg = customred},
  cell{10}{11} = {bg = customred},
  cell{11}{3,6,8,11} = {bg = customred},
  cell{13}{2-7,8,9,10} = {bg = customgreen},
  cell{14}{2-7,9,10} = {bg = customgreen},
  cell{15}{2-7,9,10} = {bg = customgreen},
  cell{16}{2-10} = {bg = customgreen},
  cell{17}{2,4,5,7,9,10} = {bg = customgreen},
  cell{18}{2-11} = {bg = customgreen},
  cell{19}{2-11} = {bg = customgreen},
  cell{20}{2-7,9,10} = {bg = customgreen},
  cell{13}{6,8,11} = {bg = customred},
  cell{14}{8,11} = {bg = customred},
  cell{15}{8,11} = {bg = customred},
  cell{16}{11} = {bg = customred},
  cell{17}{3,6,8,10,11} = {bg = customred},
  cell{20}{8,11} = {bg = customred},
  cell{1}{1} = {r=2}{},
  cell{1}{4} = {c=2}{},
  cell{1}{6} = {c=2}{},
  cell{1}{8} = {c=2}{},
  cell{3}{1} = {c=11}{},
  cell{12}{1} = {c=11}{},
  hline{1,21} = {-}{0.1em},
  hline{2,3,4,11,12,13,20} = {-}{0.05em},
  hline{2} = {2,3,4,6,8,10,11}{l},
  hline{2} = {2,3,5,7,9,10}{r},
}
{Surrogate\\Model→} & Logit \cite{zhao2021success} & SU \cite{wei2023selfuniversal} & Logit Margin \cite{weng2023logit} & & FFT \cite{zeng2024enhancing} & & CGNC \cite{fang2024clip} & & CleanSheet \cite{ge2024hijacking} & UnivIntruder \\
                    & RN50(IM)   & RN50(IM)  & RN50(IM)          & CLIP      & RN50(IM)  & CLIP      & RN50(IM)  & CLIP      & Ensemble        & CLIP      \\
\( l_{2} = 20 \)           & & & & & & & & & & \\
ResNet              & 14.4/31.5  & 25.3/47.6 & 17.1/39.0         & 5.2/15.3  & 29.5/60.5 & 9.8/23.7  & 41.3/69.6 & 6.1/15.2  & 36.0/59.5       & 51.9/80.6 \\
VGG                 & 11.6/27.3  & 23.6/44.7 & 11.4/30.2         & 2.1/7.9   & 28.7/55.1 & 5.1/15.2  & 50.1/82.3 & 25.4/43.4 & 28.6/57.7       & 55.2/85.0 \\
MobileNet           & 9.5/24.0   & 18.5/37.5 & 10.0/29.4         & 6.8/17.8  & 20.9/43.4 & 13.3/29.7 & 65.3/82.9 & 15.8/33.9 & 17.0/40.4       & 60.9/84.1 \\
ShuffleNet          & 5.2/12.9   & 6.5/16.8  & 5.0/19.4          & 8.1/19.4  & 9.0/23.6  & 14.3/29.4 & 22.0/44.7 & 8.9/25.6  & 14.3/39.6       & 50.1/83.2 \\
DenseNet            & 32.5/56.6  & 44.9/68.9 & 30.8/57.9         & 5.3/16.6  & 49.4/78.9 & 9.7/25.8  & 45.7/77.4 & 15.1/29.9 & 26.8/55.1       & 46.9/81.7 \\
ViT                 & 4.8/20.5   & 5.4/23.4  & 6.7/25.0          & 9.1/22.3  & 5.4/22.4  & 16.4/33.5 & 4.1/14.2  & 3.1/8.4   & 0.7/12.5        & 26.8/74.3 \\
Swin                & 12.1/43.0  & 18.8/65.0 & 13.4/46.7         & 4.3/12.6  & 19.5/53.4 & 7.1/19.8  & 15.9/49.8 & 1.5/6.7   & 7.2/70.7        & 30.7/93.8 \\
Average             & 12.9/30.8  & 20.4/43.4 & 13.5/35.4         & 5.8/16.0  & 23.2/48.2 & 10.8/25.3 & 34.9/60.1 & 10.8/23.3 & 18.7/47.9       & 46.1/83.2 \\
\( l_{\infty} = \frac{16}{255} \)                    & & & & & & & & & & \\
ResNet              & 7.7/20.4   & 8.9/23.6  & 7.2/18.6          & 4.7/14.7  & 26.1/45.7 & 3.4/12.0  & 39.1/63.8 & 2.8/9.7   & 13.6/28.2       & 35.0/58.1 \\
VGG                 & 3.9/12.1   & 5.4/15.4  & 3.6/9.7           & 1.6/6.8   & 19.3/37.0 & 2.0/8.2   & 42.8/75.0 & 5.6/16.8  & 12.4/23.5       & 30.6/57.8 \\
MobileNet           & 2.8/10.8   & 4.4/13.6  & 3.6/9.2           & 5.6/18.9  & 14.8/29.6 & 5.8/17.6  & 41.0/66.1 & 2.3/9.2   & 9.2/16.4        & 49.4/69.0 \\
ShuffleNet          & 0.8/4.5    & 1.0/5.4   & 0.6/2.9           & 5.5/15.5  & 4.6/14.4  & 5.9/16.5  & 11.8/31.6 & 1.8/8.0   & 6.6/8.7         & 32.9/59.2 \\
DenseNet            & 17.0/37.7  & 20.7/43.8 & 16.5/37.3         & 5.2/17.3  & 39.3/57.9 & 3.7/13.3  & 50.7/75.0 & 5.9/14.3  & 21.9/37.7       & 38.1/60.8 \\
ViT                 & 1.4/12.4   & 0.9/18.1  & 2.2/13.6          & 12.4/27.0 & 1.8/11.7  & 5.5/16.8  & 3.7/25.4  & 7.0/45.5  & 4.0/12.8        & 14.1/75.2 \\
Swin                & 3.6/27.8   & 5.7/58.1  & 5.3/29.2          & 3.5/10.8  & 9.5/35.7  & 1.7/7.5   & 10.1/61.6 & 1.6/51.8  & 7.4/33.5        & 12.7/95.9 \\
Average             & 5.3/18.0   & 6.7/25.4  & 5.6/17.2          & 5.5/15.8  & 16.5/33.2 & 4.0/13.1  & 28.5/56.9 & 3.8/22.2  & 10.7/22.9       & 30.4/68.0 \\
\end{tblr}
\vspace{-0.2cm}
\end{small}
\end{table*}

\section{Comparison with $L_2$-norm constraint.}  
\label{ap:l2norm}
To comprehensively evaluate our proposed method, we extend our analysis to include experimental results under the $L_2$-norm constraint, complementing the $L_\infty$-norm-based experiments presented in the main text. Table \ref{tab:comparison_combined} reports the Top-1 and Top-5 ASR for various adversarial attack methods, including our proposed \system{}, under an $L_2$-norm perturbation budget of $\epsilon=20$. The results demonstrate that when using CLIP as the surrogate model, \system{} exhibits superior transferability, consistently outperforming baseline methods across multiple target architectures. Notably, while some baselines achieve strong performance when employing RN50(IM) as the surrogate model, their effectiveness diminishes significantly when transferred to CLIP-based attack scenarios. In contrast, \system{} maintains high ASR, underscoring its robustness and adaptability to different surrogate models. These findings align with prior studies, such as CleanSheet \cite{ge2024hijacking}, which observed similar performance patterns under both $L_\infty$ and $L_2$ norms. The ability of \system{} to generalize effectively across both norm constraints highlights its practical utility in real-world settings.

\begin{figure}
\begin{center}
\centerline{\includegraphics[width=0.8\linewidth]{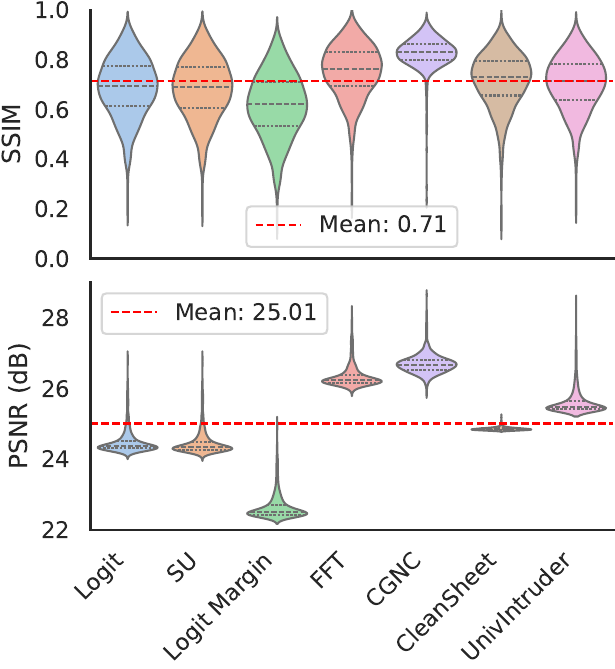}}
\vspace{-0.4cm}
\caption{\partrevisionG{PSNR and SSIM score distribution for baselines under \( l_{\infty} = \frac{16}{255} \) on ImageNet. For both metrics, higher values indicate better invisibility. Our method ranks 4th in SSIM and 3rd in PSNR out of all the 7 attacks.}}
\label{fig:psnr_ssim}
\end{center}
\vspace{-0.8cm}
\end{figure}

\section{Comparison with Lower $L_\infty$-norm constraint.}  
\label{ap:linfnorm}
We here consider a lower $L_\infty$-norm with \( l_{\infty} = \frac{16}{255} \) to evaluate a more challenging scenario for all the baselines. Table \ref{tab:comparison_combined} presents a comprehensive comparison of adversarial ASR on ImageNet under an $L_\infty$-norm constraint with a perturbation budget of $\epsilon=16/255$. Our proposed method, \system{}, uses CLIP as the surrogate model and achieves remarkable transferability across a diverse set of target models, including ResNet, VGG, MobileNet, ShuffleNet, DenseNet, ViT, and Swin. Notably, \system{} records an average Top-1 ASR of 30.4\% and Top-5 ASR of 68.0\%, outperforming most baseline methods under this stringent constraint. While methods like CGNC (RN50) achieve competitive results on specific models (e.g., 50.7\% Top-1 ASR on DenseNet), their performance is less consistent across architectures compared to \system{}. Baseline approaches such as Logit, SU, and Logit Margin, which rely on RN50(IM), exhibit significantly reduced effectiveness under this tighter perturbation limit, often falling below 20\% Top-1 ASR. In contrast, \system{} maintains robust performance, particularly on challenging models like MobileNet (49.4\% Top-1 ASR) and ShuffleNet (32.9\% Top-1 ASR), highlighting its adaptability and efficacy in generating subtle yet potent adversarial examples. This evaluation underscores the practical superiority of \system{} in constrained adversarial settings.

\end{partDsection}

\begin{partGsection}
\section{Comparison on Stealthiness Metrics.}  

Quantitative results on SSIM and PSNR show that \system{} performs not badly in stealthiness. Beyond the human evaluation in Fig. \ref{fig:human}, we assess stealthiness using objective metrics commonly employed in image editing and adversarial perturbation studies: SSIM and PSNR \cite{antsiferova2024comparing}. Higher values indicate better imperceptibility. Fig. \ref{fig:psnr_ssim} illustrates the distribution of these metrics across baseline methods under \( l_{\infty} = \frac{16}{255} \). C-GNC \cite{fang2024clip} leads with the highest SSIM and PSNR, likely due to its Gaussian smoothing, which mitigates high-frequency noise. Our method, \system{}, achieves moderate stealthiness, ranking 4th in SSIM and 3rd in PSNR out of 7 methods.

\end{partGsection}

\section{Settings in Detection Based on Robustness}  
\label{ap:robust_defense}

To evaluate robustness in feature embeddings under various transformations, we applied a range of controlled modifications to input images. The transformations used included:

\begin{itemize}
    \item \textbf{JPEG Compression}: Images were compressed with a fixed quality level of 50 to simulate lossy compression effects commonly encountered in practical scenarios.
    \item \textbf{Gaussian Blur}: A Gaussian blur filter was applied with a kernel size of 7 and a random sigma value in the range of 0.1 to 2.0, introducing controlled levels of blur.
    \item \textbf{Random Affine Transformations}: Images underwent affine transformations with random rotation up to 15 degrees, translation up to 10\% of the image dimensions, scaling within the range of 0.8 to 1.2, and a shear of up to 10 degrees.
    \item \textbf{Color Jittering}: Adjustments to brightness, contrast, saturation, and hue were applied, with variations constrained to 0.2 for brightness, contrast, and saturation, and 0.1 for hue.
    \item \textbf{Random Horizontal Flip}: Horizontal flipping was applied with a probability of 50\%.
    \item \textbf{Random Perspective Transformations}: Perspective distortions were introduced with a distortion scale of 0.5.
\end{itemize}

These transformations were used to generate perturbed versions of input images, and their corresponding feature embeddings were compared with those of the original images. This allowed us to measure the robustness of feature representations under a variety of transformations, reflecting their stability and resistance to adversarial perturbations.

\section{Experimental Setting in PubFig83}  
\label{ap:pubfig}  

\noindent \textbf{Facial Adaptation.} To improve the alignment of perturbations with individual user faces, we employ an adaptive technique that customizes perturbations for each face. Specifically, we detect 68 facial landmarks for each facial image and optimize an affine transformation matrix based on the standard landmark positions and the corresponding landmarks in the current image. This approach allows us to apply the affine transformation of the universal perturbation to each individual face, achieving better alignment of facial perturbations. This method can be described as adaptive universal adversarial perturbations.

\noindent \textbf{POOD Dataset Preparation.} We utilize CelebA \cite{liu2015celeba} as the POOD dataset. Since CelebA is annotated with 40 attributes—such as gender, hairstyle, expressions, and accessories—we treat these binary attributes as labels. To use these labels, we compile a list of attributes in textual format and input them into CLIP to generate the corresponding textual embeddings.

\begin{figure}
\begin{center}
\centerline{\includegraphics[width=1.0\linewidth]{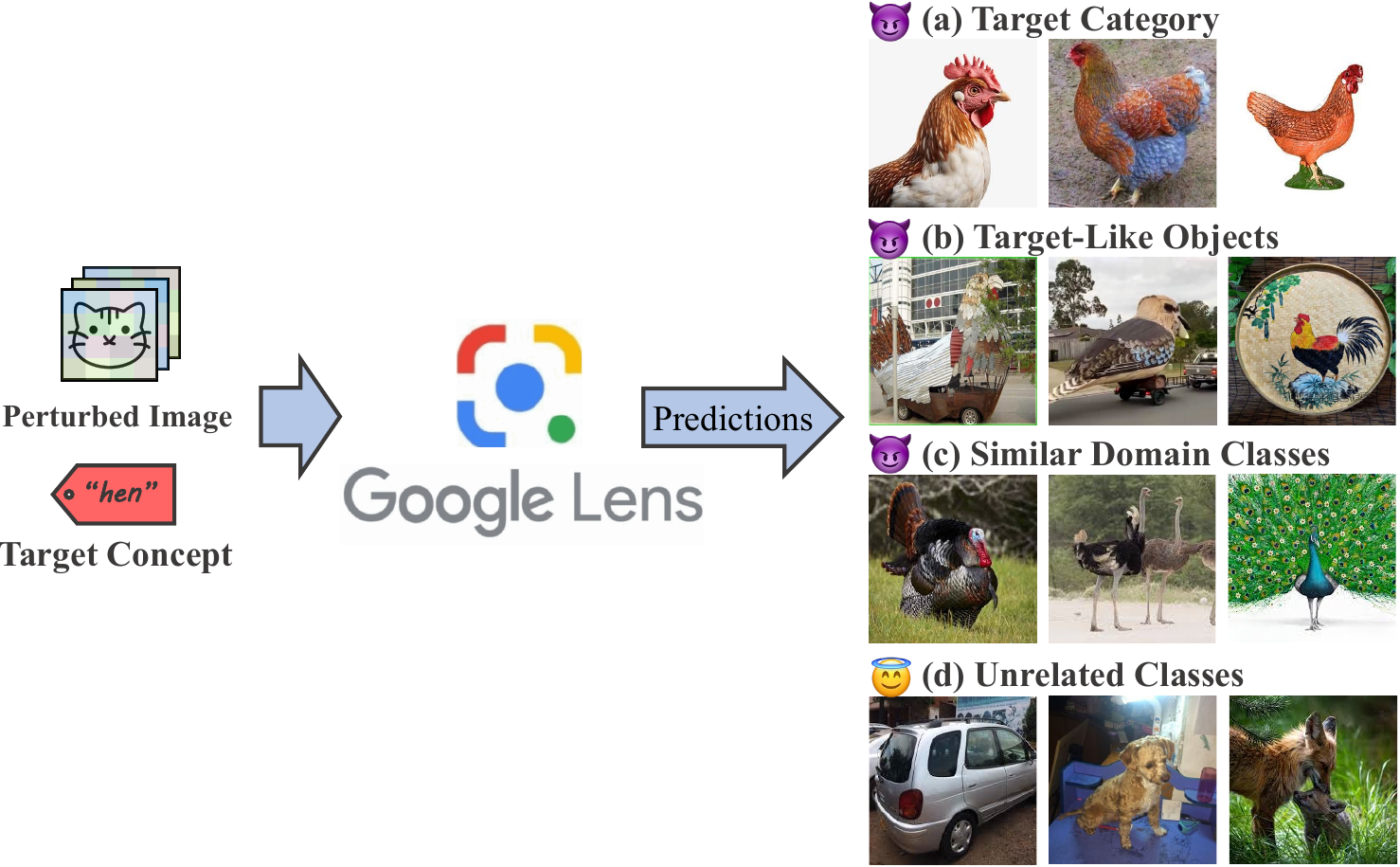}}
\vspace{-0.4cm}
\caption{Four categories of image search services and their representative sample results.}
\label{fig:metric_image_search}
\end{center}
\vspace{-0.8cm}
\end{figure}

\section{Detailed Results on CIFAR-10, CIFAR-100, Caltech-101, and ImageNet}

We present detailed experiments for CIFAR-10, CIFAR-100, Caltech-101, and ImageNet. As shown in Tables \ref{tab:cifar10}, \ref{tab:imagenet}, and \ref{tab:caltech}, our attack achieves a high ASR, demonstrating its effectiveness across datasets with varying image resolutions, from low to high.

\noindent \textbf{Reproducibility of Results.} To facilitate reproducibility, we provide open-source access to the weights used for the universal perturbations in our released code. The universal perturbations for CIFAR-10, CIFAR-100, and ImageNet datasets are made available under two perturbation budgets: (\( l_{\infty} = \frac{16}{255} \)) and (\( l_{\infty} = \frac{32}{255} \)). All tested models are sourced from reputable implementations in the machine learning community. For instance, we utilize pretrained models for CIFAR-10 and CIFAR-100 from \url{https://github.com/chenyaofo/pytorch-cifar-models}, and ImageNet-1K weights are obtained from TorchVision's official pretrained models.

\begin{table*}
\centering
\caption{The performance of \system{} on CIFAR-10 and CIFAR-100.}
\vspace{-0.4cm}
\label{tab:cifar10}
\begin{small}
\begin{tblr}{
  width = 0.9\linewidth,
  colspec = {Q[102]Q[98]Q[108]Q[173]Q[162]Q[110]Q[87]Q[87]},
  rowsep = 0.8pt,
  colsep = 0.8pt,
  cells = {c},
  cell{1}{1} = {c=8}{0.9\linewidth},
  cell{12}{1} = {c=8}{0.9\linewidth},
  hline{1,23} = {-}{0.10em},
  hline{2,5,8,11-13,16,19,22} = {-}{0.05em},
}
\textbf{CIFAR-10}   &                &                &                     &                    &                &                &                \\
Metric              & ResNet-32      & VGG-13-BN      & MobileNet V2 (0.75) & ShuffleNet V2 1.0× & RepVGG-A0      & ViT-B-16       & Swin-L         \\
CA(\%)            & 93.53          & 94.00          & 93.72               & 92.98              & 94.39          & 98.76          & 99.14          \\
ASR(\%) ↑           & 98.57          & 97.28          & 96.77               & 98.08              & 95.84          & 99.42          & 96.01          \\
Metric              & ResNet-44      & VGG-16-BN      & MobileNet V2 (1.0)  & ShuffleNet V2 1.5× & RepVGG-A1      & ViT-L-32       & Swin-V2-B      \\
CA(\%)            & 94.01          & 94.16          & 93.79               & 93.55              & 94.89          & 98.81          & 99.17          \\
ASR(\%) ↑           & 98.57          & 96.18          & 93.37               & 96.72              & 99.21          & 98.92          & 95.62          \\
Metric              & ResNet-56      & VGG-19-BN      & MobileNet V2 (1.4)  & ShuffleNet V2 2.0× & RepVGG-A2      & ViT-L-16       & Swin-V2-L      \\
CA(\%)            & 94.37          & 93.91          & 94.22               & 93.81              & 94.98          & 99.18          & 99.31          \\
ASR(\%) ↑           & 98.98          & 87.08          & 98.21               & 92.34              & 97.13          & 97.77          & 94.63          \\
\textbf{$\overline{\text{ASR}}$ (\%) ↑}  & \textbf{98.71} & \textbf{93.51} & \textbf{96.12}      & \textbf{95.71}     & \textbf{97.39} & \textbf{98.70} & \textbf{95.42} \\
\textbf{CIFAR-100 } &                &                &                     &                    &                &                &                \\
Metric              & ResNet-32      & VGG-13-BN      & MobileNet V2 (0.75) & ShuffleNet V2 1.0× & RepVGG-A0      & ViT-B-16       & Swin-L         \\
CA(\%)            & 70.16          & 74.63          & 73.61               & 72.39              & 75.22          & 90.66          & 93.57          \\
ASR(\%) ↑           & 95.98          & 96.75          & 98.11               & 86.70              & 93.71          & 97.00          & 94.84          \\
Metric              & ResNet-44      & VGG-16-BN      & MobileNet V2 (1.0)  & ShuffleNet V2 1.5× & RepVGG-A1      & ViT-L-32       & Swin-V2-B      \\
CA(\%)            & 71.63          & 74.00          & 74.20               & 73.91              & 76.12          & 91.69          & 93.59          \\
ASR(\%) ↑           & 95.57          & 94.60          & 93.49               & 84.98              & 96.60          & 98.19          & 96.94          \\
Metric              & ResNet-56      & VGG-19-BN      & MobileNet V2 (1.4)  & ShuffleNet V2 2.0× & RepVGG-A2      & ViT-L-16       & Swin-V2-L      \\
CA(\%)            & 72.63          & 73.87          & 75.98               & 75.35              & 77.18          & 93.72          & 94.42          \\
ASR(\%) ↑           & 94.68          & 92.53          & 98.09               & 87.96              & 94.10          & 96.29          & 97.81          \\
\textbf{$\overline{\text{ASR}}$ (\%) ↑}  & \textbf{95.41} & \textbf{94.63} & \textbf{96.56}      & \textbf{86.55}     & \textbf{94.80} & \textbf{97.16} & \textbf{96.53} 
\end{tblr}
\caption*{
\scalebox{1.0}{\small Note that all the pre-trained target models except ViT and Swin are from \url{https://github.com/chenyaofo/pytorch-cifar-models}.
}}
\vspace{-0.6cm}
\end{small}
\end{table*}

\begin{table*}
\centering
\caption{The performance of \system{} on ImageNet.}
\vspace{-0.4cm}
\label{tab:imagenet}
\begin{small}

\begin{tblr}{
  width = 0.9\linewidth,
  rowsep = 0.8pt,
  colsep = 0.8pt,
  colspec = {Q[92]Q[67]Q[67]Q[67]Q[67]Q[67]Q[67]Q[60]Q[60]Q[65]Q[65]Q[83]Q[83]},
  cells = {c},
  cell{1}{1} = {c=13}{0.9\linewidth},
  cell{2}{1} = {r=2}{},
  cell{2}{2} = {c=2}{0.134\linewidth},
  cell{2}{4} = {c=2}{0.134\linewidth},
  cell{2}{6} = {c=2}{0.134\linewidth},
  cell{2}{8} = {c=2}{0.12\linewidth},
  cell{2}{10} = {c=2}{0.13\linewidth},
  cell{2}{12} = {c=2}{0.166\linewidth},
  cell{6}{1} = {r=2}{},
  cell{6}{2} = {c=2}{0.134\linewidth},
  cell{6}{4} = {c=2}{0.134\linewidth},
  cell{6}{6} = {c=2}{0.134\linewidth},
  cell{6}{8} = {c=2}{0.12\linewidth},
  cell{6}{10} = {c=2}{0.13\linewidth},
  cell{6}{12} = {c=2}{0.166\linewidth},
  cell{10}{1} = {r=2}{},
  cell{10}{2} = {c=2}{0.134\linewidth},
  cell{10}{4} = {c=2}{0.134\linewidth},
  cell{10}{6} = {c=2}{0.134\linewidth},
  cell{10}{8} = {c=2}{0.12\linewidth},
  cell{10}{10} = {c=2}{0.13\linewidth},
  cell{10}{12} = {c=2}{0.166\linewidth},
  cell{14}{1} = {r=2}{},
  cell{14}{2} = {c=2}{0.134\linewidth},
  cell{14}{4} = {c=2}{0.134\linewidth},
  cell{14}{6} = {c=2}{0.134\linewidth},
  cell{14}{8} = {c=2}{0.12\linewidth},
  cell{14}{10} = {c=2}{0.13\linewidth},
  cell{14}{12} = {c=2}{0.166\linewidth},
  cell{16}{12} = {c=2}{0.166\linewidth},
  hline{1,18} = {-}{0.10em},
  hline{2,6,10,14} = {-}{0.05em},
  }
\textbf{ImageNet} &              &       &              &       &              &       &            &       &              &       &                    &                \\
Metric            & ResNet-18    &       & ResNet-34    &       & ResNet-50    &       & ResNet-101 &       & WRN-50-2     &       & WRN-101-2          &                \\
                  & top-1        & top-5 & top-1        & top-5 & top-1        & top-5 & top-1      & top-5 & top-1        & top-5 & top-1              & top-5          \\
CA (\%)         & 69.76        & 89.08 & 73.31        & 91.42 & 76.13        & 92.86 & 77.37      & 93.55 & 78.47        & 94.09 & 78.85              & 94.28          \\
ASR(\%) ↑         & 70.58        & 95.77 & 71.75        & 95.56 & 83.39        & 95.97 & 84.69      & 95.21 & 77.30        & 94.62 & 75.36              & 94.16          \\
Metric            & VGG-11       &       & VGG-11-BN    &       & VGG-16       &       & VGG-16-BN  &       & MobileNet V2 &       & ShuffleNet V2 1.0× &                \\
                  & top-1        & top-5 & top-1        & top-5 & top-1        & top-5 & top-1      & top-5 & top-1        & top-5 & top-1              & top-5          \\
CA (\%)         & 69.02        & 88.63 & 70.37        & 89.81 & 71.59        & 90.38 & 73.36      & 91.52 & 71.88        & 90.29 & 69.36              & 88.32          \\
ASR(\%) ↑         & 74.08        & 92.79 & 65.65        & 93.11 & 69.15        & 90.51 & 54.38      & 92.09 & 85.15        & 97.49 & 75.39              & 93.64          \\
Metric            & DenseNet-121 &       & DenseNet-169 &       & DenseNet-201 &       & ViT-B-32   &       & ViT-B-16     &       & ViT-L-32           &                \\
                  & top-1        & top-5 & top-1        & top-5 & top-1        & top-5 & top-1      & top-5 & top-1        & top-5 & top-1              & top-5          \\
CA (\%)         & 74.43        & 91.97 & 75.60        & 92.81 & 76.90        & 93.37 & 75.91      & 92.47 & 81.07        & 95.32 & 76.97              & 93.07          \\
ASR(\%) ↑         & 78.38        & 95.27 & 83.14        & 95.46 & 78.80        & 96.01 & 61.88      & 89.77 & 69.68        & 99.29 & 62.64              & 91.51          \\
Metric            & ViT-L-16     &       & Swin-T       &       & Swin-S       &       & Swin-V2-T  &       & Swin-V2-S    &       & \textbf{Average}   &                \\
                  & top-1        & top-5 & top-1        & top-5 & top-1        & top-5 & top-1      & top-5 & top-1        & top-5 & \textbf{top-1}     & \textbf{top-5} \\
CA (\%)         & 79.66        & 94.64 & 81.47        & 95.78 & 83.20        & 96.36 & 82.07      & 96.13 & 83.71        & 96.82 & \textbf{-}         &                \\
ASR(\%) ↑         & 59.59        & 97.95 & 52.03        & 99.74 & 27.54        & 99.65 & 51.95      & 99.54 & 31.72        & 99.74 & \textbf{67.14}     & \textbf{95.43} 
\end{tblr}
\caption*{
\scalebox{1.0}{\small Note that all the pre-trained target models are from Torchvision.
}}
\end{small}
\vspace{-0.6cm}
\end{table*}

\begin{table}[H]
\centering
\caption{The results of \system{} on Caltech-101.}
\vspace{-0.4cm}
\label{tab:caltech}
\begin{small}

\begin{tblr}{
  width = \linewidth,
  rowsep = 0.8pt,
  colsep = 0.8pt,
  colspec = {Q[185]Q[265]Q[219]Q[260]},
  cells = {c},
  cell{1}{1} = {c=4}{0.921\linewidth},
  hline{1,26} = {-}{0.10em},
  hline{2,5,8,11,14,17,20,23} = {-}{0.05em},
}
\textbf{Caltech101} &                       &                     &                       \\
\textbf{Metric}     & \textbf{ResNet-18}    & \textbf{ResNet-101} & \textbf{VGG-11-BN}    \\
CA (\%)           & 91.42                 & 93.32               & 89.63                 \\
ASR(\%) ↑           & 89.64                 & 85.99               & 84.08                 \\
\textbf{Metric}     & \textbf{ResNet-50}    & \textbf{VGG-11}     & \textbf{VGG-16-BN}    \\
CA (\%)           & 93.89                 & 88.77               & 91.36                 \\
ASR(\%) ↑           & 84.77                 & 87.03               & 91.55                 \\
\textbf{Metric}     & \textbf{ResNet-34}    & \textbf{VGG-16}     & \textbf{MobileNet V2} \\
CA (\%)           & 93.15                 & 89.06               & 90.73                 \\
ASR(\%) ↑           & 86.39                 & 85.87               & 92.13                 \\
\textbf{Metric}     & \textbf{WRN-50-2}    & \textbf{WRN-101-2}     & \textbf{ShuffleNet V2 1.0×} \\
CA (\%)           & 93.26                 & 94.41               & 85.20                 \\
ASR(\%) ↑           & 73.65                 & 71.63               & 72.73                 \\
\textbf{Metric}     & \textbf{DenseNet-169} & \textbf{ViT-b-32}   & \textbf{ViT-l-16}     \\
CA (\%)           & 93.55                 & 94.64               & 96.82                 \\
ASR(\%) ↑           & 85.58                 & 90.74               & 87.97                 \\
\textbf{Metric}     & \textbf{DenseNet-121} & \textbf{ViT-l-32}   & \textbf{Swin-v2-t}    \\
CA (\%)           & 92.40                 & 93.95               & 97.05                 \\
ASR(\%) ↑           & 87.38                 & 89.15               & 87.05                 \\
\textbf{Metric}     & \textbf{DenseNet-201} & \textbf{ViT-b-16}   & \textbf{Swin-v2-s}    \\
CA (\%)           & 94.53                 & 96.66               & 97.26                 \\
ASR(\%) ↑           & 86.51                 & 91.16               & 85.48                 \\
\textbf{Metric}     & \textbf{Swin-t}       & \textbf{Swin-s}     & \textbf{Average}      \\
CA (\%)           & 97.00                 & 97.18               & \textbf{93.27}            \\
ASR(\%) ↑           & 88.05                 & 89.15               & \textbf{85.81}        
\end{tblr}
\end{small}
\vspace{-0.3cm}
\end{table}

\begin{table*}[ht]
\centering
\caption{\partrevisionE{Comparison of our attack against robust models with baseline methods on ImageNet.}}
\vspace{-0.4cm}
\label{tab:robust_imagenet}
\begin{small}
{
\begin{tabular}{lcccccccccccc}
\toprule
\multirow{2}{*}{\textbf{Method}} & \multirow{2}{*}{\textbf{Model}} & \multirow{2}{*}{\textbf{CA (\%)}} & \multicolumn{2}{c}{\textbf{FFT \cite{zeng2024enhancing}}} & \multicolumn{2}{c}{\textbf{Logit Margin \cite{weng2023logit}}} & \multicolumn{2}{c}{\textbf{CGNC \cite{fang2024clip}}} & \multicolumn{2}{c}{\textbf{CleanSheet \cite{ge2024hijacking}}} & \multicolumn{2}{c}{\textbf{UnivIntruder}} \\
\cmidrule(lr){4-5} \cmidrule(lr){6-7} \cmidrule(lr){8-9} \cmidrule(lr){10-11} \cmidrule(lr){12-13}
& & & \textbf{Top-1} & \textbf{Top-5} & \textbf{Top-1} & \textbf{Top-5} & \textbf{Top-1} & \textbf{Top-5} & \textbf{Top-1} & \textbf{Top-5} & \textbf{Top-1} & \textbf{Top-5} \\
\midrule
\cite{amini2024meansparse} & CNX-L & 78.0 & 0.3 & 1.7 & 0 & 0.5 & 2.6 & 7.6 & 0.2 & 4.6 & 5.9 & 20.3 \\
\cite{bai2024mixednuts} & CNX-V2-L+Swin-L & 81.5 & 9.5 & 32.9 & 0.6 & 1.6 & 13.2 & 45.2 & 2.3 & 9.8 & 7.5 & 47.2 \\
\cite{liu2024comprehensive} & Swin-B & 76.2 & 0.1 & 0.7 & 0 & 0.1 & 1.4 & 3.6 & 1.1 & 3.7 & 7.3 & 30.7 \\
\cite{liu2024comprehensive} & Swin-L & 78.9 & 0.1 & 1.3 & 0 & 0.2 & 1.5 & 3.9 & 0.1 & 2.2 & 7.0 & 39.2 \\
\cite{mo2022adversarial} & Swin-B & 74.7 & 0.2 & 1.8 & 0 & 0.1 & 0.2 & 0.9 & 0.1 & 2.3 & 7.4 & 14.2 \\
\cite{singh2024revisiting} & CNX-L+ConvStem & 77.0 & 0 & 0 & 0 & 0 & 0 & 0 & 1.7 & 3.4 & 5.9 & 18.2 \\
\cite{singh2024revisiting} & ViT-B+ConvStem & 76.3 & 0.1 & 0.5 & 0 & 0 & 0.1 & 0.6 & 0 & 1.1 & 12.2 & 40.5 \\
\cite{singh2024revisiting} & ViT-S+ConvStem & 72.6 & 0 & 0.1 & 0 & 0 & 0.2 & 1.6 & 0 & 1 & 7.1 & 26.9 \\
\bottomrule
\end{tabular}
}
\end{small}
\vspace{-0.3cm}
\end{table*}

\section{Metric Design for Image Searching Services}

\label{ap:metric_image}

As shown in Fig. \ref{fig:metric_image_search}, we categorize all possible predictions from image search services into four different classes:

\begin{enumerate}
    \item \textbf{Target Category}: This encompasses all recognized results that map clearly to the \textit{target concept}. For example, if our target is ``hen,'' this category would include not just ``hen'' but any synonyms or subspecies directly related to the target.
    
    \item \textbf{Target-Like Objects}: This category includes results where the system recognizes an object that \textit{shares shapes, textures, or colors} with the target concept, but is still labeled as some other object. For instance, a plate or purse might be recognized as having the ``hen'' pattern, thus mixing features of the target concept with those of the actual object.
    
    \item \textbf{Similar Domain Classes}: These are conceptually or visually related classes that are \textit{similar to the target domain} but not exactly the target. For instance, if the target is a ``hen'', other bird species fall into this category.
    
    \item \textbf{Unrelated Classes}: This category includes predictions for the \textit{original class} of the image (if it remains recognizable) or any other third-party class that is neither the target nor visually/domain-related to it. These indicate failed attacks.
\end{enumerate}

Following this classification, we define two metrics. (1) \textit{Special-ASR}: The proportion of predictions labeled as the Target Category (i.e., those that precisely match the targeted concept). (2) \textit{General-ASR}: The proportion of predictions labeled as Target Category, Target-Like Objects, or Similar Domain Classes combined.

\section{More results in attack targets}
In Table \ref{tab:target_cifar10}, we analyzed the performance across various target classes on CIFAR-10. Here, we extend our analysis to target classes on CIFAR-100 and ImageNet. The results, shown in Table \ref{tab:target_cifar100} and Table \ref{tab:target_imagenet}, confirm that our method performs effectively across most classes, aligning with our observations on CIFAR-10. However, the ASR for certain challenging classes, such as class 68 on CIFAR-100 and class 208 on ImageNet, is relatively low. This is because some target classes rely on specific robust features (like colors or some low-frequency features) for classification that are difficult to replicate with adversarial examples, as previously discussed in \cite{fang2024clip}.

\begin{table}[H]
\centering
\caption{ASR of different target classes on CIFAR-100.}
\vspace{-0.4cm}
\label{tab:target_cifar100}
\begin{small}
\begin{tblr}{
  width = \linewidth,
  rowsep = 0.8pt,
  colsep = 0.8pt,
  colspec = {Q[127]Q[127]Q[127]Q[127]Q[127]Q[127]},
  cells = {c},
  hline{1,10} = {-}{0.10em},
  hline{2,9} = {-}{0.05em},
}
\textbf{Target Class→} & \textbf{8 (bicycle)} & \textbf{28 (cup)} & \textbf{48 (motorcycle)} & \textbf{68 (road)} & \textbf{88 (tiger)} \\
ResNet      & 95.3      & 65.3      & 82.4      & 14.6      & 87.0      \\
VGG         & 90.9      & 75.8      & 87.1      & 26.5      & 80.5      \\
MobileNet   & 95.6      & 79.9      & 83.9      & 15.5      & 91.5      \\
ShuffleNet  & 85.0      & 63.7      & 80.5      & 21.2      & 83.6      \\
RepVGG      & 94.8      & 71.7      & 88.7      & 25.6      & 90.4      \\
ViT         & 97.4      & 97.9      & 90.2      & 93.1      & 95.2      \\
Swin        & 96.2      & 96.3      & 92.5      & 87.1      & 95.2      \\
Average     & 93.6      & 78.7      & 86.5      & 40.5      & 89.1      
\end{tblr}
\end{small}
\vspace{-0.1cm}
\end{table}

\begin{table}[H]
\centering
\caption{ASR of different target classes on ImageNet. Each cell shows \textit{top-1 ASR / top-5 ASR}.}
\vspace{-0.4cm}
\label{tab:target_imagenet}
\begin{small}
\begin{tblr}{
  width = \linewidth,
  rowsep = 0.8pt,
  colsep = 0.8pt,
  colspec = {Q[127]Q[127]Q[127]Q[127]Q[127]Q[127]},
  cells = {c},
  hline{1,10} = {-}{0.10em},
  hline{2,9} = {-}{0.05em},
}
\textbf{Target Class→} & \textbf{8 (hen)} & \textbf{208 (labrador retriever)} & \textbf{408 (amphibious vehicle)} & \textbf{608 (jeans)} & \textbf{808 (sombrero)} \\
ResNet      & 77.2/95.2   & 38.8/78.2   & 75.1/89.1   & 75.4/87.8   & 79.2/90.4   \\
VGG         & 65.8/92.1   & 17.1/38.6    & 72.0/89.4   & 85.2/92.5   & 83.0/93.1   \\
MobileNet   & 85.1/97.5   & 24.9/75.9   & 72.8/86.5   & 74.1/87.2   & 71.4/87.1   \\
ShuffleNet  & 75.4/93.6   & 20.1/50.3   & 63.3/83.7   & 41.5/62.8   & 56.7/77.6   \\
DenseNet    & 80.1/95.6   & 22.9/75.9   & 80.2/93.2   & 88.9/95.8   & 88.2/95.1   \\
ViT         & 63.4/94.6   & 18.6/58.7    & 39.5/80.0   & 44.5/84.3   & 59.0/88.4   \\
Swin        & 40.8/99.7   & 18.3/66.2    & 47.7/97.7   & 59.6/99.3   & 45.8/97.6   \\
Average     & 69.7/95.5   & 23.0/63.4   & 64.4/88.5   & 67.0/87.1   & 69.1/89.9   
\end{tblr}
\end{small}
\vspace{-0.1cm}
\end{table}

\section{More Results on Robust Models}
In Table \ref{tab:adversarial_training}, we previously discussed the performance of four different types of robust models on CIFAR-10. In this section, we extend our analysis to robust models on CIFAR-100 and ImageNet. Due to the limited availability of open-source robust models for these two datasets compared to CIFAR-10, we present the results method by method, without categorizing them.

The results in Table \ref{tab:robust_cifar100} and Table \ref{tab:robust_imagenet} demonstrate that \partrevisionE{various defense methods, including FFT, Logit Margin, CGNC, and CleanSheet, are effective in mitigating our attack, consistent with our observations on CIFAR-10. However, our method, UnivIntruder, shows higher ASR compared to these defense methods, indicating its effectiveness in challenging robust models.} However, the challenges of high training resource consumption and reduced clean accuracy persist.

\begin{table}[H]
\centering
\caption{Our attack against robust models on CIFAR-100.}
\vspace{-0.4cm}
\label{tab:robust_cifar100}
\begin{small}
\begin{tabular}{lcccc}
\toprule
\textbf{Method} & \textbf{Model} & \textbf{CA (\%)} & \textbf{ASR (\%)} & \textbf{RA (\%)} \\
\midrule
\cite{bai2024improving}&Ensemble-EDM  & 85.2 & 36.5 & 39.8 \\
\cite{bai2024improving}&Ensemble-Trades  & 80.2 & 37.1 & 38.1 \\
\cite{bai2024mixednuts}&Ensemble-Mixed  & 83.1 & 44.7 & 38.5 \\
\cite{cui2021learnable}&WRN-34-10  & 70.3 & 34.1 & 28.8 \\
\cite{cui2023decoupled}&WRN-28-10  & 73.9 & 25.0 & 40.3 \\
\cite{cui2023decoupled}&WRN-34-10  & 65.9 & 16.2 & 43.5 \\
\cite{gowal2020uncovering}&WRN-70-16  & 69.2 & 26.0 & 40.1 \\
\cite{rebuffi2021fixing}&WRN-70-16  & 63.6 & 17.7 & 38.9 \\
\cite{wang2023better}&WRN-28-10  & 72.6 & 19.3 & 44.2 \\
\cite{wang2023better}&WRN-70-16  & 75.2 & 29.8 & 41.3 \\
\bottomrule
\end{tabular}
\end{small}
\end{table}

\section{More Case Study Results}
\begin{figure*}
\begin{center}
\centerline{\includegraphics[width=1.0\linewidth]{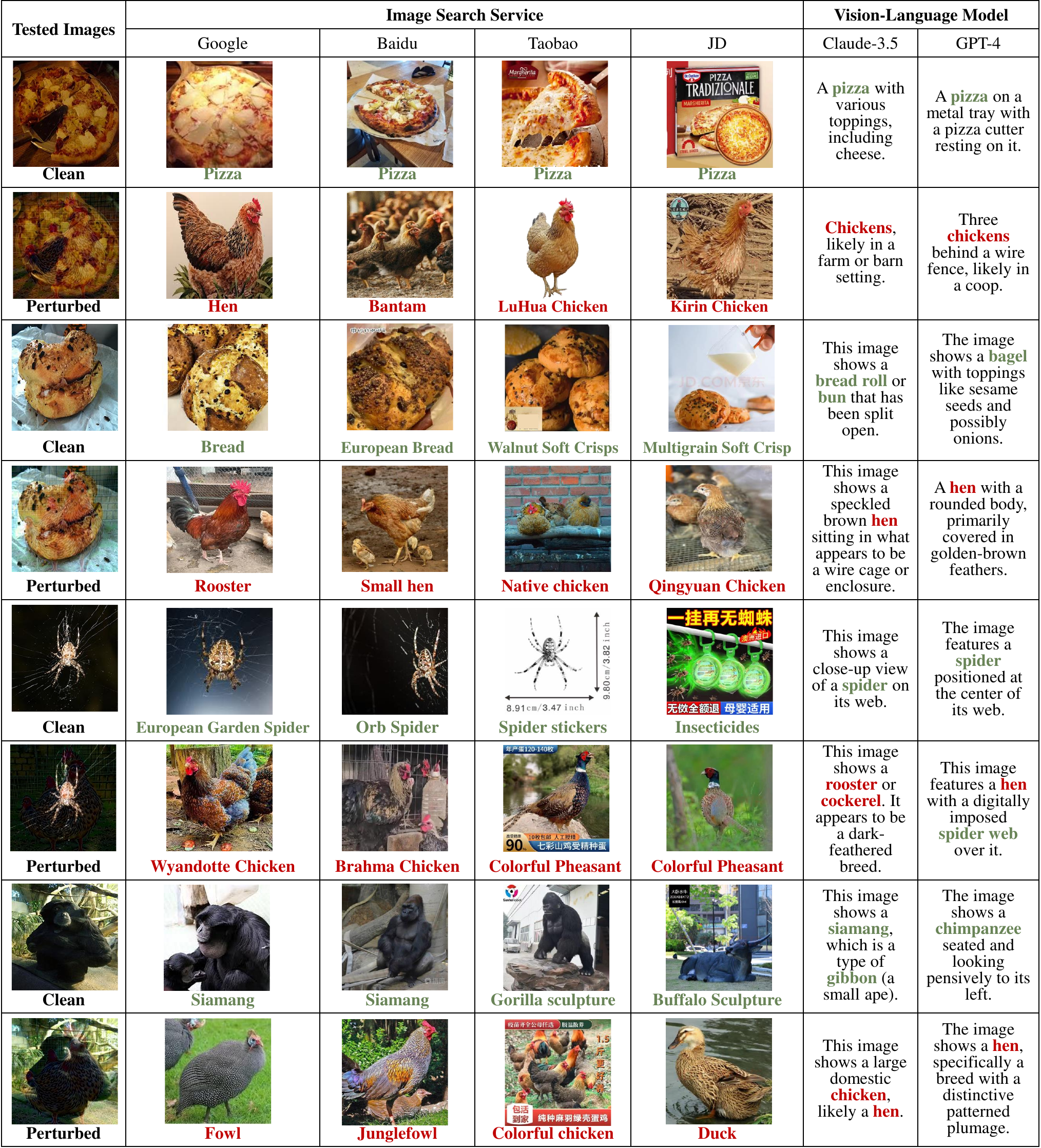}}
\caption{More Case studies. Due to wildlife protection, the spider and the siamang are not recognized on shopping sites like Taobao and JD. This experiment uses the default prompt \textit{``Please concisely classify what is in this image''} for LLM evaluation. Feel free to screenshot the perturbed image (ensure at least 256 resolution) to verify.}
\label{fig:case_study_appendix}
\end{center}
\end{figure*}

We present additional visualization results of the generated perturbations and results on real applications in Figure \ref{fig:case_study_appendix}. All images are selected from the ImageNet validation set. The results show that our perturbations effectively mislead inputs towards the target concept, regardless of the original image features. Notably, both image search services and large vision language models are compromised in this manner, highlighting the high generalizability of our attack.

\end{document}